\documentclass[%
reprint,
superscriptaddress,
 amsmath,amssymb,
pra,
]{revtex4-1}
\usepackage{graphicx}
\usepackage{dcolumn}
\usepackage{bm}
\usepackage{color}

\begin{document}

\preprint{???}

\title{
High-power laser experiment forming a supercritical collisionless shock
in a magnetized uniform plasma at rest
}
%


\author{R.~Yamazaki}
\affiliation{Department of Physical Sciences, Aoyama Gakuin University, 5-10-1 Fuchinobe, Sagamihara, Kanagawa 252-5258, Japan.}
\email[]{ryo@phys.aoyama.ac.jp (R.Y.)}
\affiliation{Institute of Laser Engineering, Osaka University, 2-6, Yamadaoka, Suita, Osaka 565-0871, Japan.}

\author{S. Matsukiyo}
\affiliation{Faculty of Engineering Sciences, Kyushu University, 6-1 Kasuga-Koen, Kasuga, Fukuoka 816-8580, Japan.}

\author{T.~Morita}
\affiliation{Faculty of Engineering Sciences, Kyushu University, 6-1 Kasuga-Koen, Kasuga, Fukuoka 816-8580, Japan.}

\author{S. J. Tanaka}
\affiliation{Department of Physical Sciences, Aoyama Gakuin University, 5-10-1 Fuchinobe, Sagamihara, Kanagawa 252-5258, Japan.}

\author{T.~Umeda}
\affiliation{Institute for Space-Earth Environmental Research, Nagoya University, Furo-cho, Chikusa, Nagoya, Aichi 464-8602, Japan.}


\author{K.~Aihara}
\affiliation{Department of Physical Sciences, Aoyama Gakuin University, 5-10-1 Fuchinobe, Sagamihara, Kanagawa 252-5258, Japan.}

\author{M.~Edamoto}
\affiliation{Interdisciplinary Graduate School of Engineering Sciences, Kyushu University, Kasuga 816-8580, Japan.}

\author{S.~Egashira}
\affiliation{Graduate School of Science, Osaka University, 1-1 Machikane-yama, Toyonaka, Osaka 560-0043, Japan.}

\author{R.~Hatsuyama}
\affiliation{Interdisciplinary Graduate School of Engineering Sciences, Kyushu University, Kasuga 816-8580, Japan.}

\author{T.~Higuchi}
\affiliation{Interdisciplinary Graduate School of Engineering Sciences, Kyushu University, Kasuga 816-8580, Japan.}

\author{T.~Hihara}
\affiliation{Graduate School of Engineering, Osaka University, 2-1, Yamadaoka, Suita, Osaka 565-0871, Japan.}

\author{Y.~Horie}
\affiliation{Interdisciplinary Graduate School of Engineering Sciences, Kyushu University, Kasuga 816-8580, Japan.}

\author{M.~Hoshino}
\affiliation{Department of Earth and Planetary Science, The University of Tokyo, 7-3-1 Hongo, Bunkyo, Tokyo 113-0033, Japan.}

\author{A.~Ishii}
\affiliation{Max Planck Institute for Gravitational Physics (Albert Einstein Institute), Am M\"{u}hlenberg 1, Potsdam-Golm, 14476, Germany}

\author{N.~Ishizaka}
\affiliation{Department of Physical Sciences, Aoyama Gakuin University, 5-10-1 Fuchinobe, Sagamihara, Kanagawa 252-5258, Japan.}

\author{Y.~Itadani}
\affiliation{Interdisciplinary Graduate School of Engineering Sciences, Kyushu University, Kasuga 816-8580, Japan.}

\author{T.~Izumi}
\affiliation{Graduate School of Science, Osaka University, 1-1 Machikane-yama, Toyonaka, Osaka 560-0043, Japan.}

\author{S.~Kambayashi}
\affiliation{Department of Physical Sciences, Aoyama Gakuin University, 5-10-1 Fuchinobe, Sagamihara, Kanagawa 252-5258, Japan.}

\author{S.~Kakuchi}
\affiliation{Department of Physical Sciences, Aoyama Gakuin University, 5-10-1 Fuchinobe, Sagamihara, Kanagawa 252-5258, Japan.}

\author{N.~Katsuki}
\affiliation{Interdisciplinary Graduate School of Engineering Sciences, Kyushu University, Kasuga 816-8580, Japan.}

\author{R.~Kawamura}
\affiliation{Department of Physical Sciences, Aoyama Gakuin University, 5-10-1 Fuchinobe, Sagamihara, Kanagawa 252-5258, Japan.}

\author{Y.~Kawamura}
\affiliation{Department of Physical Sciences, Aoyama Gakuin University, 5-10-1 Fuchinobe, Sagamihara, Kanagawa 252-5258, Japan.}

\author{S.~Kisaka}
\affiliation{Department of Physical Science, Hiroshima University, Higashi-Hiroshima 739-8526, Japan.}

\author{T.~Kojima}
\affiliation{Interdisciplinary Graduate School of Engineering Sciences, Kyushu University, Kasuga 816-8580, Japan.}

\author{A.~Konuma}
\affiliation{Institute for Laser Science, The University of Electro-Communications, 1-5-1 Chofugaoka, Chofu, Tokyo 182-8585, Japan.}

\author{R.~Kumar}
\affiliation{Graduate School of Science, Osaka University, 1-1 Machikane-yama, Toyonaka, Osaka 560-0043, Japan.}

\author{T.~Minami}
\affiliation{Graduate School of Engineering, Osaka University, 2-1, Yamadaoka, Suita, Osaka 565-0871, Japan.}

\author{I.~Miyata}
\affiliation{Department of Physical Sciences, Aoyama Gakuin University, 5-10-1 Fuchinobe, Sagamihara, Kanagawa 252-5258, Japan.}

\author{T.~Moritaka}
\affiliation{Fundamental Physics Simulation Research Division, National Institute for Fusion Science, 322-6 Oroshi-cho, Toki, 509-5292, Japan.}

\author{Y.~Murakami}
\affiliation{Interdisciplinary Graduate School of Engineering Sciences, Kyushu University, Kasuga 816-8580, Japan.}

\author{K.~Nagashima}
\affiliation{Interdisciplinary Graduate School of Engineering Sciences, Kyushu University, Kasuga 816-8580, Japan.}

\author{Y.~Nakagawa}
\affiliation{Graduate School of Science, Osaka University, 1-1 Machikane-yama, Toyonaka, Osaka 560-0043, Japan.}

\author{T.~Nishimoto}
\affiliation{School of Engineering, Osaka University, 2-1, Yamadaoka, Suita, Osaka 565-0871, Japan.}

\author{Y.~Nishioka}
\affiliation{Interdisciplinary Graduate School of Engineering Sciences, Kyushu University, Kasuga 816-8580, Japan.}

\author{Y.~Ohira}
\affiliation{Department of Earth and Planetary Science, The University of Tokyo, 7-3-1 Hongo, Bunkyo, Tokyo 113-0033, Japan.}

\author{N.~Ohnishi}
\affiliation{Department of Aerospace Engineering, Tohoku University,  6-6 Aramaki Aza Aoba, Aoba, Sendai, Miyagi 980-8579, Japan.}

\author{M.~Ota}
\affiliation{Graduate School of Science, Osaka University, 1-1 Machikane-yama, Toyonaka, Osaka 560-0043, Japan.}

\author{N.~Ozaki}
\affiliation{Graduate School of Engineering, Osaka University, 2-1, Yamadaoka, Suita, Osaka 565-0871, Japan.}

\author{T.~Sano}
\affiliation{Institute of Laser Engineering, Osaka University, 2-6, Yamadaoka, Suita, Osaka 565-0871, Japan.}

\author{K.~Sakai}
\affiliation{Graduate School of Engineering, Osaka University, 2-1, Yamadaoka, Suita, Osaka 565-0871, Japan.}

\author{S.~Sei}
\affiliation{Department of Physical Sciences, Aoyama Gakuin University, 5-10-1 Fuchinobe, Sagamihara, Kanagawa 252-5258, Japan.}

\author{J.~Shiota}
\affiliation{Department of Physical Sciences, Aoyama Gakuin University, 5-10-1 Fuchinobe, Sagamihara, Kanagawa 252-5258, Japan.}

\author{Y.~Shoji}
\affiliation{Department of Physical Sciences, Aoyama Gakuin University, 5-10-1 Fuchinobe, Sagamihara, Kanagawa 252-5258, Japan.}

\author{K.~Sugiyama}
\affiliation{Department of Physical Sciences, Aoyama Gakuin University, 5-10-1 Fuchinobe, Sagamihara, Kanagawa 252-5258, Japan.}

\author{D.~Suzuki}
\affiliation{Department of Physical Sciences, Aoyama Gakuin University, 5-10-1 Fuchinobe, Sagamihara, Kanagawa 252-5258, Japan.}

\author{M.~Takagi}
\affiliation{Interdisciplinary Graduate School of Engineering Sciences, Kyushu University, Kasuga 816-8580, Japan.}

\author{H.~Toda}
\affiliation{Department of Physical Sciences, Aoyama Gakuin University, 5-10-1 Fuchinobe, Sagamihara, Kanagawa 252-5258, Japan.}

\author{S.~Tomita}
\affiliation{Astronomical Institute, Tohoku University,  6-3 Aramaki, Aoba-ku, Sendai, Miyagi 980-8578, Japan.}
\affiliation{Frontier Research Institute for Interdisciplinary Sciences, Tohoku University, Sendai, 980-8578, Japan}

\author{S.~Tomiya}
\affiliation{Department of Physical Sciences, Aoyama Gakuin University, 5-10-1 Fuchinobe, Sagamihara, Kanagawa 252-5258, Japan.}

\author{H.~Yoneda}
\affiliation{Institute for Laser Science, The University of Electro-Communications, 1-5-1 Chofugaoka, Chofu, Tokyo 182-8585, Japan.}


\author{T.~Takezaki}
\affiliation{Department of Creative Engineering, National Institute of Technology, Kitakyushu College, 5-20-1 Shii, Kokuraminamiku, Kitakyushu, Fukuoka, 802-0985, Japan.}
\affiliation{Faculty of Engineering, University of Toyama, 3190, Gofuku, Toyama 930-8555, Japan.}

\author{K.~Tomita}
\affiliation{Faculty of Engineering Sciences, Kyushu University, 6-1 Kasuga-Koen, Kasuga, Fukuoka 816-8580, Japan.}
\affiliation{Division of Quantum Science and Engineering, Hokkaido University, Sapporo 060-8628, Japan.}

\author{Y.~Kuramitsu}
\affiliation{Graduate School of Engineering, Osaka University, 2-1, Yamadaoka, Suita, Osaka 565-0871, Japan.}

\author{Y.~Sakawa}
\affiliation{Institute of Laser Engineering, Osaka University, 2-6, Yamadaoka, Suita, Osaka 565-0871, Japan.}




\date{\today}

\begin{abstract}

We present a new experimental method to generate 
quasi-perpendicular supercritical magnetized collisionless shocks.
In our experiment, 
ambient nitrogen (N) plasma is at rest and 
well-magnetized, and it has uniform mass density.
The plasma is pushed
by laser-driven ablation aluminum (Al) plasma.
Streaked optical pyrometry and spatially resolved laser collective Thomson scattering
clarify structures of plasma density and temperatures, 
which are compared with one-dimensional particle-in-cell simulations.
It is indicated that 
just after the laser irradiation, the Al plasma 
is magnetized by a self-generated Biermann battery field, and 
the plasma slaps the incident N plasma.
The compressed external field in the N plasma reflects N ions, leading to counter-streaming magnetized N flows.
Namely we identify the edge of the reflected N ions.
Such interacting plasmas form a magnetized collisionless shock.
\end{abstract}

\pacs{Valid PACS appear here}
\maketitle


\section{INTRODUCTION}\label{sec:intro}

\

Collisionless shocks are ubiquitous in astrophysical objects like supernova remnants and
solar-terrestrial and laboratory plasmas 
\cite{Balogh2013,Burgess2015}.
When the upstream low-entropy flow comes into the shock, the kinetic energy is converted
into various forms like high-temperature ions and electrons, magnetic turbulence, and nonthermal particles.
However, despite state-of-the-art observations
\cite{Bamba2003,Johlander2016},
particle-in-cell (PIC) simulations \cite{Scholer2003,Lembege2009,Matsukiyo2011,Matsumoto2013,Sironi2013,Umeda2014,Yamazaki2019},
and analytical arguments \cite{Vink2014},
the detailed mechanism of the energy dissipation is not fully understood.
In many cases, the upstream plasma is magnetized, and the
pre-existing and/or self-generated magnetic fields around the shock
 work as a  ``catalyst'' in the process of kinetic energy dissipation.

The laboratory experiment using high-power lasers is another method to study collisionless shocks.
Laser-produced plasma is fast-moving and long-lived.
Therefore, it has been expected that large-scale, long-time evolution of the plasma interaction can be seen,
which is unachievable by current PIC simulations.
There have been experiments to excite several kinds of collisionless shocks:
electrostatic shocks \cite{Kuramitsu2011,Morita2010,Sakawa2016,Yuan2017}, 
and Weibel-mediated shocks \cite{Ross2017,Li2019,Fiuza2020}, 
as well as subcritical \cite{Morita2013,Niemann2014,Schaeffer2015} and
supercritical \cite{Kuramitsu2016,Schaeffer2017,Schaeffer2019} magnetized shocks.
Previous experiments have revealed that even in the unmagnetized case, 
a self-generated magnetic field
 (i.e., Biermann field) is crucial in the ion reflection \cite{Huntington2015}, which plays an important role in the formation of 
perpendicular shocks.
This is also indicated by one-dimensional (1D) PIC simulation \cite{Umeda2019}.

In many astrophysical magnetized collisionless shocks producing cosmic rays,
they are supercritical 
(Alfv\'{e}n Mach number $M_A\gtrsim3$)
at which a part of incoming ions are reflected upstream
and gyrate back into the shock front \cite{Sckopke1990}, causing two-stream instabilities to generate plasma waves
which lead to the particle scattering and acceleration.
So far, due to limited space and time, 
no experimental results of clear formation of such a supercritical shock have been reported,
although some authors claimed the observation of a precursor which was expected to evolve into the shock
if the plasma interaction proceeds 
\cite{Li2019,Schaeffer2017,Schaeffer2019}.
In this paper, we report our experiments to generate supercritical magnetized collisionless shocks,
which is compared with 1D PIC simulations.
Unlike previous experiments \cite{Schaeffer2017,Schaeffer2019},
our method can make 
a fully magnetized plasma at rest with uniform mass density \cite{Shoji2016},
so that upstream plasma parameters are well determined.

\section{Experimental Setup}\label{sec:setup}

\begin{figure}[t]
\includegraphics[width=70mm, bb= 0 0 900 450]{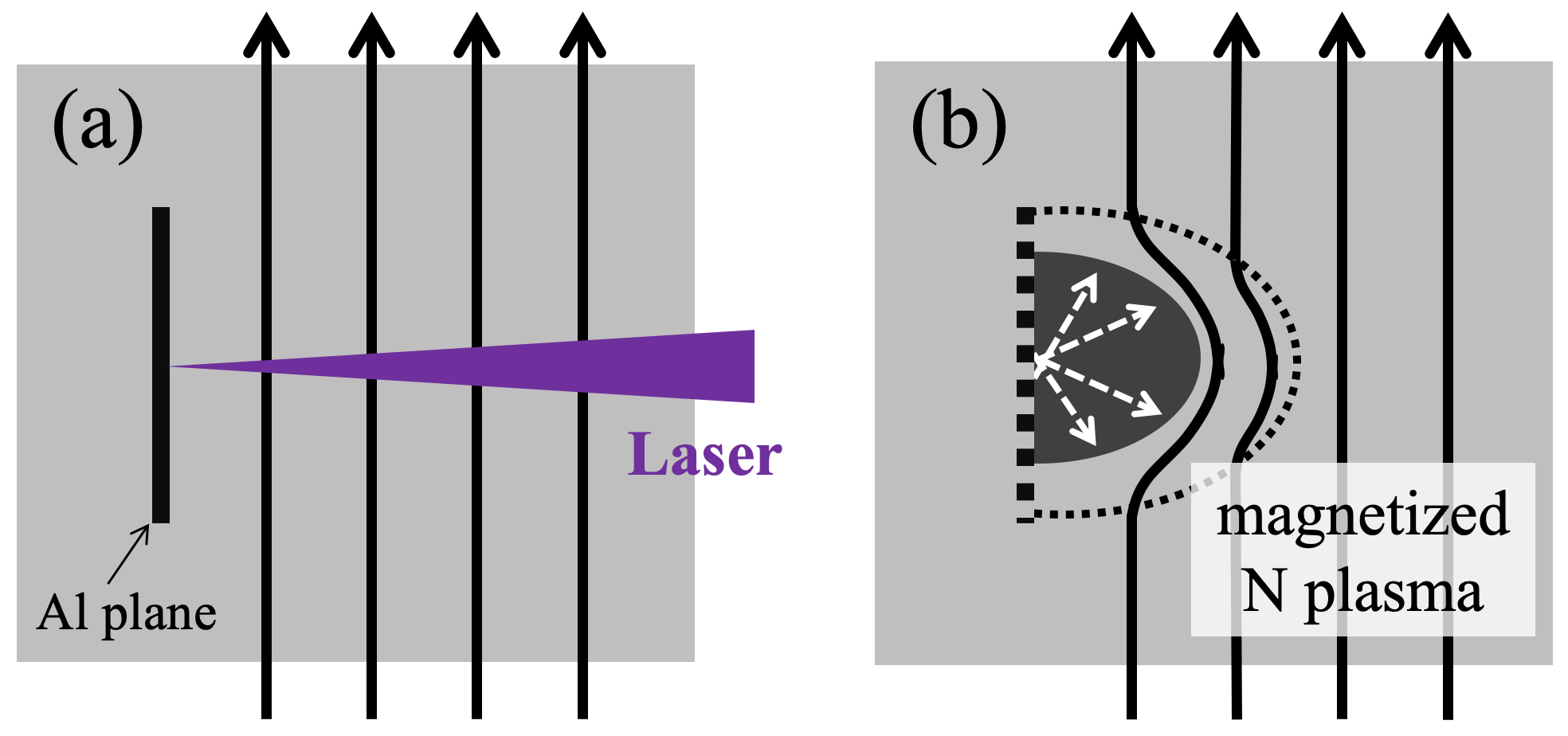}
\caption{
Schematic view of our experiment (a) before and (b) after the shot.
Solid arrows represent an external magnetic field.
Al plasma expands (white arrows) and pushes magnetized N plasma  to generate a collisionless shock (dotted curve).
\label{fig:setup}
}
\end{figure}

We used Gekko-XII HIPER Laser system (wavelength 1053~nm, pulse duration 1.3~ns,
energy 690~J~per beam, focal spot size 2.8~mm).
An aluminum (Al) plane target with thickness 2~mm was irradiated by 
four beams simultaneously, resulting laser intensity
of $3.4\times10^{13}$W~cm$^{-2}$ on the target.
A schematic side view of the target and laser configuration is shown in Fig.~\ref{fig:setup}.
Before the laser shot, the chamber was filled with ambient nitrogen (N) gas with pressure $P_N=5$~Torr.
Just before the  shot, the external magnetic field ($B_0=3.6$~T) was applied. 
The ambient gas was ionized by ionizing photons from Al plasma,
becoming magnetized plasma with N ion density $n_N=3.2\times10^{17}$cm$^{-3}$.
Subsequently, Al plasma (dark gray region in Fig.~\ref{fig:setup}(b)) pushed the magnetized N plasma to generate a magnetized collisionless shock
(dotted curve in Fig.~\ref{fig:setup}(b)).

\begin{figure}[t]
\centering
\includegraphics[width=110mm, bb=0 0 1000 570]{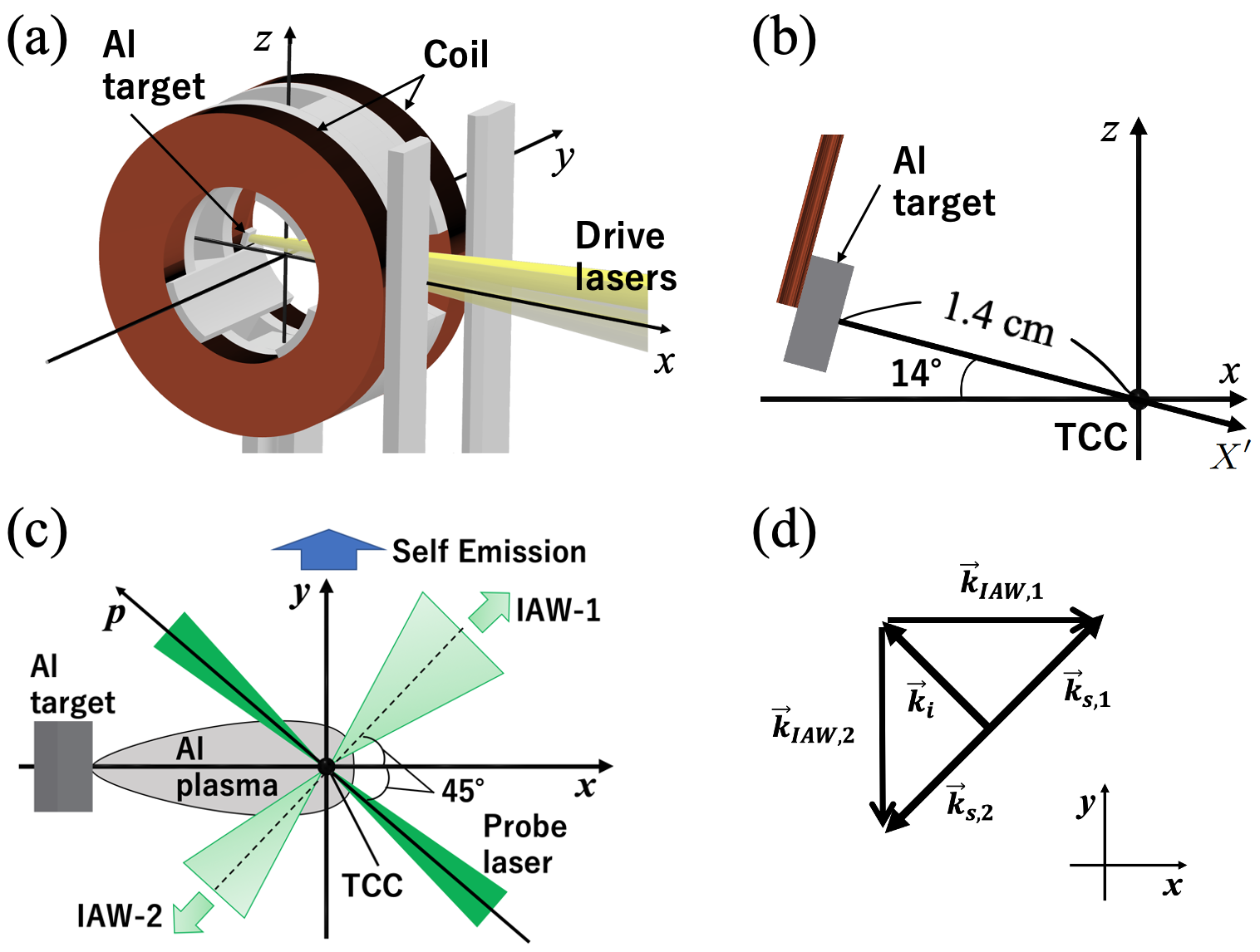}
\caption{
(a) Bird's eye view of  experimental setup.
(b) The side view of the setup. The target normal ($X'$-axis) is in the $x$-$z$ plane.
(c) The top view of the setup. The probe laser ($\vec{k}_i$: $p$-axis) for measurements of TS ion feature (IAW)
focuses at TCC and the 
scattered lights are measured from two different directions IAW-1 ($\vec{k}_{s,1}$) 
and IAW-2 ($\vec{k}_{s,2}$), and (d) the measurement wave numbers
$\vec{k}_{\rm IAW,1}=\vec{k}_{s,1}-\vec{k}_i$ and $\vec{k}_{\rm IAW,2}=\vec{k}_{s,1}-\vec{k}_i$ are roughly
longitudinal and transverse to the flow, respectively.
\label{fig:setup2}
}
\end{figure}

As shown in Fig.~\ref{fig:setup2},
the $z$-axis is the vertical, and the central axis of coils is along the $y$-axis.
The target chamber center (TCC) is located at $(x,y,z)=(0,0,0)$.
Separation between the target surface and TCC was 1.4~cm, and target normal was
$\hat{n}=(\cos14^\circ,0,-\sin14^\circ)$.
The external magnetic field was applied using an electromagnetic coil
consisting of four 50-turn coils connected in parallel.
The inner and outer diameters of the coils were 60 and 110~mm, respectively,
and two of them were placed at $y=\pm25$~mm,
to generate almost uniform magnetic field perpendicular to the plasma expansion
direction as shown in 
Figs.~\ref{fig:setup2}(a) and \ref{fig:setup2}(c).
This electromagnetic coil was driven by a small pulse-powered
circuit \cite{Edamoto18} consisting of four capacitors ($4\times1.5$~mF),
each charged with a voltage of 1.4~kV,
resulting in a quasi-static current of 5.3~kA in each coil and a uniform magnetic field
of 3.6~T inside the coils.
The time duration of the field of approximately 100~$\mu$s is sufficiently larger than
the typical timescale of the plasma propagation,
and this field is quasistatic during the plasma expansion.

Using streaked optical pyrometry (SOP),
the time evolution of the plasma self-emission 
(at a wavelength of 450~nm)
along the target normal ($X'$-axis) was observed from the $y$ direction.
The plasmas were also diagnosed with collective laser Thomson scattering (TS) method
\cite{sheffield2010plasma}.
A probe laser (Nd:YAG, wavelength 532~nm, energy 370~mJ in $\sim10$~ns) with wave number $\vec{k}_i$
went through the plasma in the horizontal plane, $z=0$, at an angle $45^\circ$ from the $x$ and $y$-axis
($p$-axis: see Fig.~\ref{fig:setup2}(c)).
The scattered light with wave number $\vec{k}_s$ was detected from two directions both of which are 
$90^\circ$ from the incident direction.
As a result, one of the measurement wave number, $\vec{k}=\vec{k}_s-\vec{k}_i$, is toward the $x$-axis,
and the other toward the $-y$ direction (Fig.~\ref{fig:setup2}(d)).
Triple grating spectrometers were used to achieve a good spectral resolution of $\approx10$ and $\approx18$~pm for
IAW-1 and IAW-2, respectively \cite{Tomita2017,Morita2019}.
We recorded the scattered light of ion feature
with an intensified charge-couple device (ICCD) with 3~ns exposure time.
In this paper, we discuss the role of self-generated field (via Biermann battery effect) 
in Al plasma \cite{Umeda2019}.
Unfortunately, there was no direct measurement of magnetic fields in this experiment 
to conclusively indicate that the Biermann field is playing an important role.

Before this experiment with N gas, we had performed similar experiments but with different ambient gas, hydrogen and helium. 
This was because it had been expected that the ion gyro radius and period could be small if the gas were fully ionized, which would help us make the field of view
 of our plasma measurement smaller. 
 The use of a simple gas would make physical interpretation clear. 
 Using Gekko-XII HIPER Laser system, we had various shots with different total laser energy and intensity, 
 changing the number of beams and/or focal spot size. 
 However, our TS measurements could not identify hydrogen or helium plasma at rest
 in the upstream region sufficiently before the Al piston plasma arrived. 
 Hence, we concluded that photoionization of hydrogen or helium gas was difficult for our experimental setup, 
at least for lasers like Gekko-XII HIPER lasers. 
The reason is that the number of ionizing photons for hydrogen or helium gas is too small. 
For example, photoionization cross section for hydrogen atom takes maximum at the photon 
absorption edge ($=13.6$~eV), and above this photon energy, the cross section approximately scales 
as $E_{ph}{}^{-3}$, where $E_{ph}$ is the photon energy.
Typical photon energy from target plasma just after the shot is $E_{ph}\sim$~keV in our laser intensity range, 
hence the photoionization cross section becomes very small, $\sim10^{-23}$cm$^2$.
Then,  the mean-free path of photons with energy $E_{ph}\sim$~keV is $\sim10^5(n_g/10^{18}{\rm cm}^{-3})^{-1}$cm, 
where $n_g$ is the hydrogen gas density. Our system size is 1--10~cm, 
so that only a small fraction of keV photons ionize the hydrogen atoms, resulting in a very small ionization fraction. 
On the other hand, since the absorption edge is much higher for nitrogen ($\approx400$~eV, depending on charge states of N ions), 
the photoionization cross section 
for the nitrogen atom is much larger ($\sim10^{-19}$cm$^2$ for $E_{ph}\sim$~keV), which makes the upstream plasma generation much easier.

\section{Analysis of experimental data}\label{sec:results}

\subsection{Analysis of plasma self-emission}

\subsubsection{Case of $P_N=B_0=0$}

\begin{figure*}[!hbt]
\centering
\includegraphics[width=860mm, bb=0 0 5700 420]{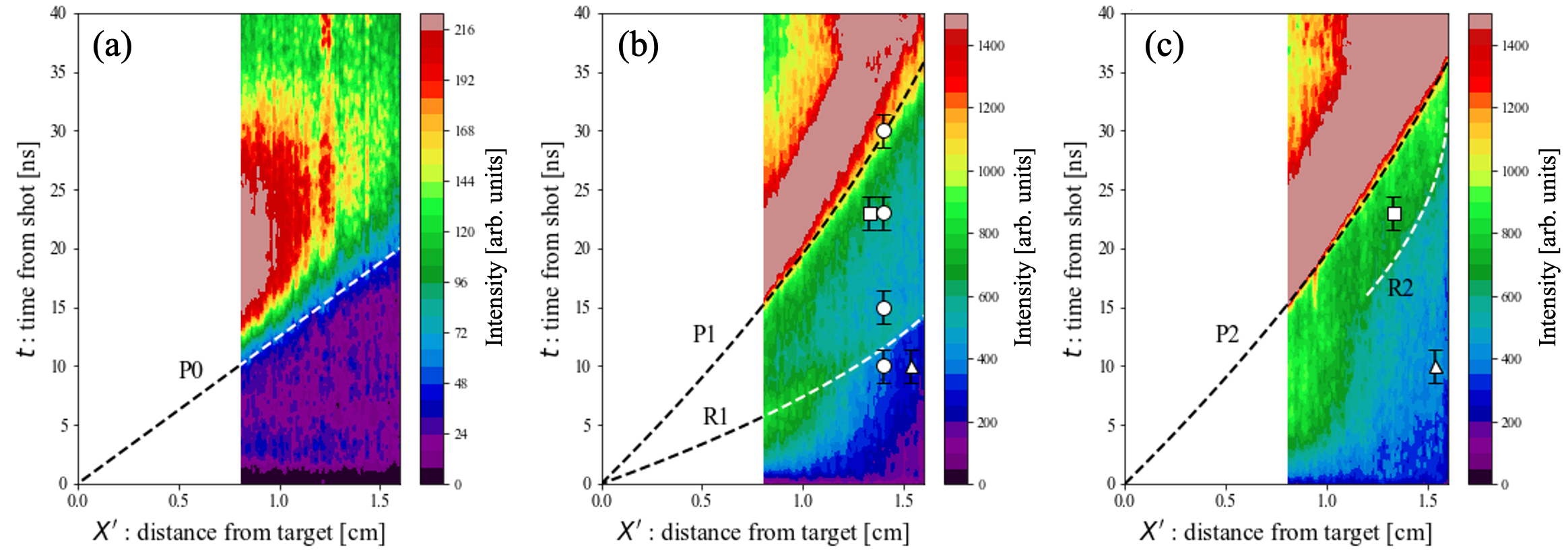}
\caption{
SOP images for cases with
(a) no ambient gas ($P_N=0$) and no external magnetic field ($B_0=0$),
(b) $P_N=5$~Torr and $B_0=0$,
and (c) $P_N=5$~Torr and $B_0=3.6$~T.
The dashed line P0 in panel (a) shows a constant velocity of 800~km~s$^{-1}$.
The curve of constant deceleration R1 in panel (b) is described by parameters
$v_0=1600$~km~s$^{-1}$ and 
$t_0=48$~ns, and
curves P1 and P2 are represented by
$v_0=590$~km~s$^{-1}$ and $t_0=148$~ns.
TCC is located at $X'=1.4$~cm ($p=0$), and epochs of TS measurements 
shown in Fig.~\ref{fig:LTS_B=0} are shown
by  white circles with error bars meaning gate width.
Assuming a plane wave with normal vector along the $X'$ direction, we also put
white squares representing positions and times of TS measurements in Fig.~\ref{fig:extmag}(b) ($p=1$~mm),
and white triangles where we estimate upstream plasma parameters ($p=-2$~mm).
\label{fig:sop}
}
\end{figure*}

First, we show the SOP result (Fig.~\ref{fig:sop}(a)) of a shot without ambient gas and external magnetic field
($P_N=B_0=0$)
to clarify the properties of piston (Al) plasma.
The Al plasma weakly emits light and freely expands with density decreasing with time.

\begin{figure*}[!hbt]
\centering
\includegraphics[width=680mm, bb=0 0 4500 380]{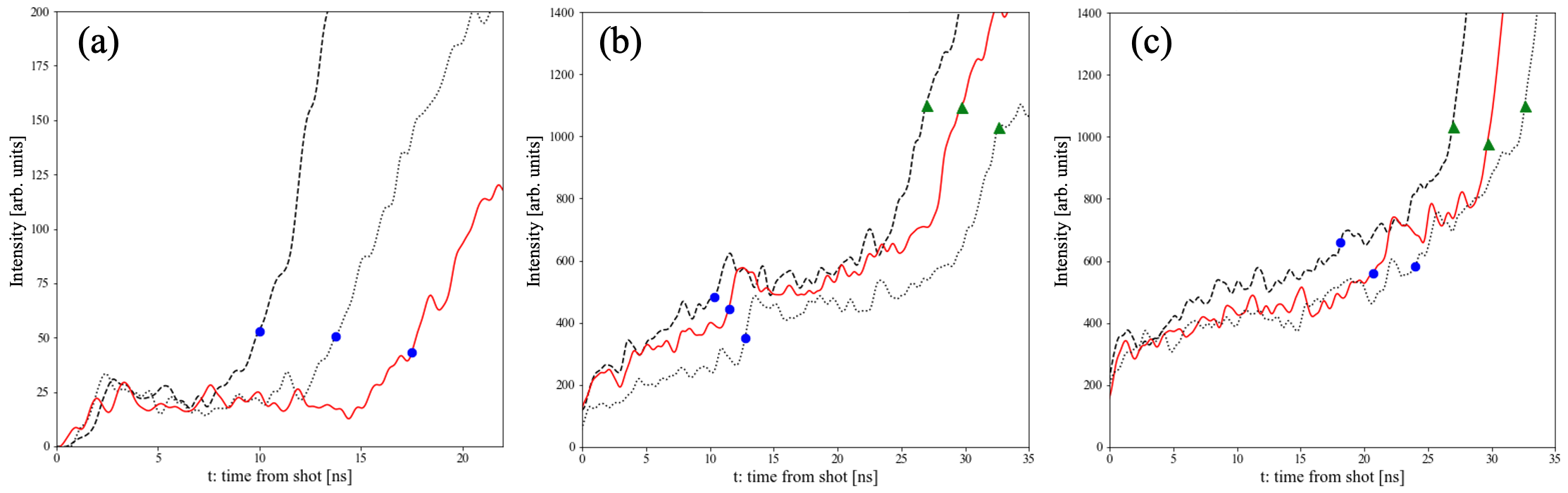}
\caption{
Time evolution of self-emission intensity (the intensity as a function of time) at fixed positions $X'$,
 for cases with
(a) no ambient gas ($P_N=0$) and no external magnetic field ($B_0=0$),
(b) $P_N=5$~Torr and $B_0=0$,
and (c) $P_N=5$~Torr and $B_0=3.6$~T.
In panel~(a), the black-dashed, black-dotted, and red-solid curves show
temporal evolutions at $X'=0.8$, 1.1, and 1.4~cm (TCC), respectively.
Blue circles in panel~(a) on each curve represent the time of passage at each position 
of a trajectory 
P0 in the $X'-t$ plane (Fig.~\ref{fig:sop}(a)) represented by 
$X'=v_{\rm Al,0} t$  with a constant velocity $v_{\rm Al,0}=800$~km~s$^{-1}$.
In panels~(b) and (c), black-dashed, red-solid, and black-dotted lines are for
$X'=1.3$, 1.4 (TCC), and 1.5~cm, respectively.
Blue circles in panels (b) and (c) represent the same as panel (a) but for R1 and R2, respectively,
whose functional form is given by Eq.~(\ref{eq:X_of_t}).
Similarly, green triangles in panels~(b) and (c) represent the time of passage of P1 and P2, respectively.
\label{fig:sop_TE_SE}
}
\end{figure*}

We show in Fig.~\ref{fig:sop_TE_SE}(a) the time evolution of SOP counts at fixed positions $X'=0.8$, 1.1, and 1.4~cm (TCC).
After the shot, the background intensity is on average $\approx25$ in our unit, and it is variable because of statistical fluctuation.
This may come from stray light of HIPER lasers, probe laser for TS measurement, and streak detector noise.
After a while, the intensity starts to increase when Al plasma arrives;
for example, at $X'=0.8$~cm (black-dashed line in Fig.~\ref{fig:sop_TE_SE}(a)), the intensity becomes twice the background level ($\approx50$)
at $t\approx10$~ns.
Assuming that the Al plasma is freely expanding,
we estimate the head speed of the Al plasma $v_{\rm Al,0}\approx0.8$~cm$/10$~ns~$=800$~km~s$^{-1}$
(see blue circles in Fig.~\ref{fig:sop_TE_SE}(a)).
The line  ``P0'' in Fig.~\ref{fig:sop}(a) represents
$X'=v_{\rm Al,0} t$ with a constant velocity $v_{\rm Al,0}=800$~km~s$^{-1}$.

\subsubsection{Case of $P_N= 5$~Torr and $B_0=0$}

Second, we had a shot with ambient N pressure  $P_N= 5$~Torr but without external magnetic field ($B_0=0$).
The result of SOP is shown in Fig.~\ref{fig:sop}(b).
The interaction between Al and N plasmas made much brighter emission than in the case of $P_N= 0$.
The edge of the brightest part   (denoted by ``P1'' in Fig.~\ref{fig:sop}(b)) propagates
with a speed $\sim350$~km~s$^{-1}$ at $t=20$~ns.
The more rapid structure ``R1'' goes ahead of P1 with a velocity of $\sim700$~km~s$^{-1}$ at $t=15$~ns.
If P1 and R1 arise at the target ($X'=0$) at $t=0$, they cannot be explained by constant-velocity motions.
Instead, assuming a quadratic function of time,
we determined their trajectories as described below.

We obtained the functional form, $X'(t)$, of P1
as in the following.
First, we made  spatial profiles of the self-emission intensity (for fixed time) every 1~ns from 15 to 35~ns.
Second, we found that the regions with SOP count of $\approx1000$ (yellow region in Fig.~\ref{fig:sop}(b))
have large intensity gradient,
whose scale length is $\Delta x\lesssim0.3$~mm.
Hence for each time, we found the value of $X'$ coordinate at which SOP count equals 1000, and we regard
the points $(X',t)$ as the position of P1 at each epoch.
Third, we adopt a quadratic function of time,
\begin{equation}
X'(t)=v_{0}t(1-t/t_0)~~,
\label{eq:X_of_t}
\end{equation}
where constants $v_0$ and $t_0$ are initial velocity and a break time, respectively.
Using this functional form, the points $(X',t)$ are fitted with least-squares method to get
$v_0=591\pm6$~km~s$^{-1}$ and $t_0=150\pm7$~ns.
Here, error means statistical in the fitting. 
Soon we will apply the same method to similar edge structure  ``P2'' in the case of 
 $P_N= 5$~Torr and $B_0=3.6$~T (Fig.~\ref{fig:sop}(c)), and derive similar values of 
 $v_0$ and $t_0$.
They coincide with each other within errors,
so that we adopt common values $v_0=590$~km~s$^{-1}$ and $t_0=148$~ns in 
dashed lines for P1 and P2
in Figs.~\ref{fig:sop}(b) and \ref{fig:sop}(c), respectively.

For the structure R1, we could not apply the same method as for P1,
because the SOP count variation (that is, the density gradient) around R1 is much less than P1.
Scale length of the gradient of R1 is $\Delta x\approx2$~mm,
so that it is difficult to define  the positions of R1.
Hence fitting by eye, we determined parameter values of $v_0$ and $t_0$.
Fortunately,
one can see from Fig.~\ref{fig:sop}(b) a clear boundary around $X'\approx1.2$--1.5~cm 
for $t\approx8$--12~ns. 
In Fig.~\ref{fig:sop}(b),
we draw the dashed line   as R1 along with this boundary,
assuming the functional form given by Eq.~(1) with $v_0=1600$~km~s$^{-1}$ and 
$t_0=48$~ns.
It can be confirmed from Fig.~\ref{fig:sop_TE_SE}(b), showing the time evolution of the intensity at fixed positions 
$X'=1.3$, 1.4, and 1.5~cm, 
that R1 is in the period of abrupt intensity increase.

As seen in Fig.~\ref{fig:sop}(b), self-emission intensity is not uniform upstream of R1.
This indicates the ionization of upstream N plasma is not uniform
(see \S~\ref{app:SOPinhomo} for detailed discussion).

\subsubsection{Case of $P_N= 5$~Torr and $B_0=3.6$~T}

\begin{figure}[t]
\centering
\includegraphics[width=110mm, bb= 0 0 1200 770]{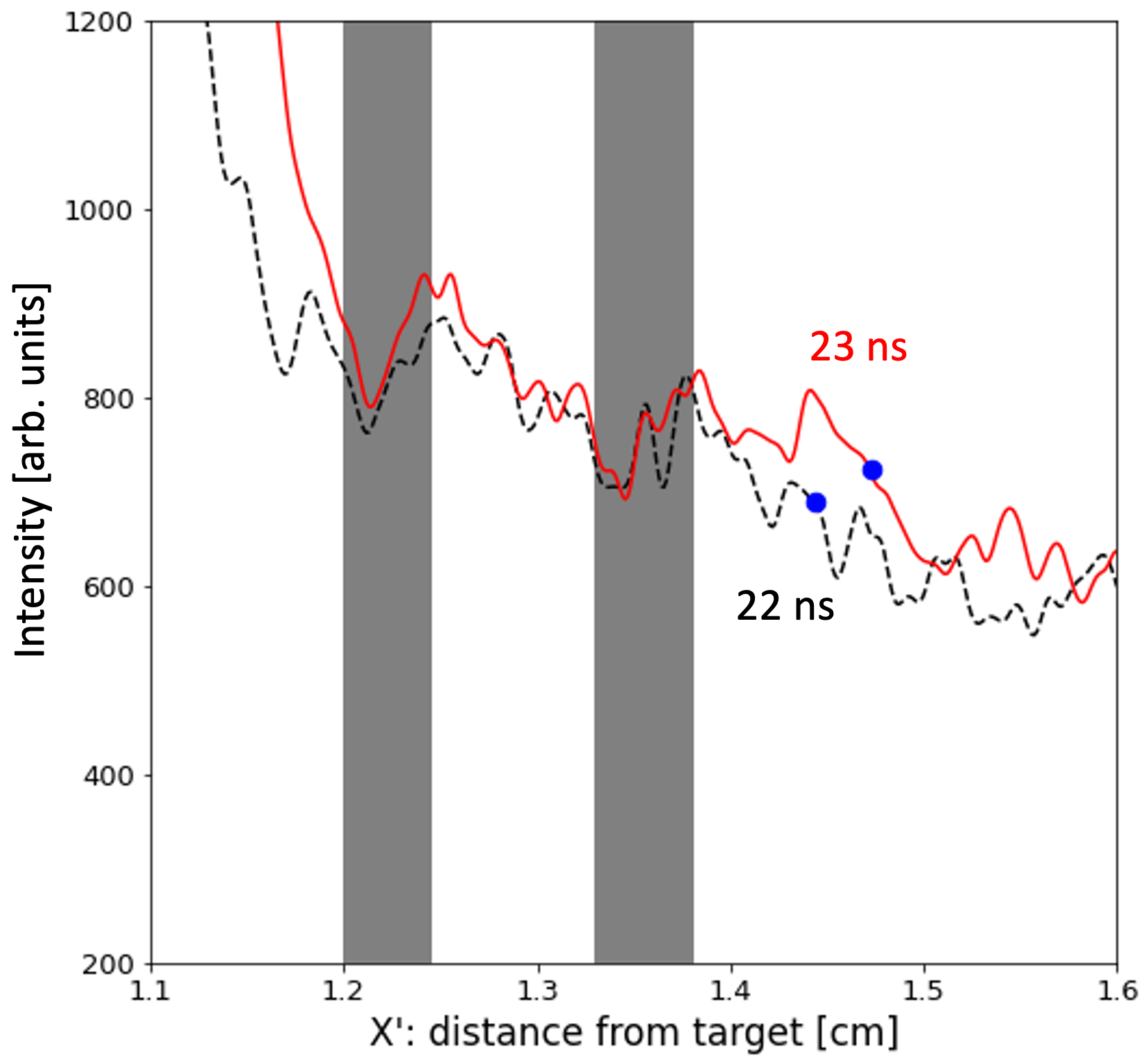}
\caption{
Spatial distribution of self-emission for the case of $P_N=5$~Torr and $B_0=3.6$~T at 
$t=22$~ns (black-dashed line) and 23~ns (red-solid line),
that is, one-dimensional slices of 
Fig.~\ref{fig:sop}(c)
at these epochs.
In the gray-shaded regions, the detector gain becomes smaller due to pixel damage.
Blue circles on each curve represent the position of passage at each epoch
of a trajectory R2 represented by Eq.~(\ref{eq:X_of_t}) with
constants $v_0=1000$~km~s$^{-1}$ and $t_0=64$~ns. 
\label{fig:selfemi23nsB}
}
\end{figure}

Third, we performed a shot with 
$P_N= 5$~Torr and $B_0=3.6$~T.
Clear difference of SOP results between with and without external field cases
can be seen (Figs.~\ref{fig:sop}(b) and \ref{fig:sop}(c)).
A small jump (denoted by ``R2'' ) is identified at 
$X'\approx1.47$~cm at $t=23$~ns at which we had a TS measurement
(see also Fig.~\ref{fig:selfemi23nsB}).
The location of the edge of the brightest region, P2, is almost the same as P1 at any time.
Separation between R2 and P2 is smaller than that between R1 and P1 in the case of $B_0=0$,
suggesting that ion dynamics is changed by the external  field.

As already described previously, 
we determined the trajectory of the structure P2 as in the same manner for P1.
It is found that P2 is explained by the functional form
given by Eq.~(\ref{eq:X_of_t}) with fitted values $v_0=592\pm5$~km~s$^{-1}$ and $t_0=148\pm5$~ns,
which  coincide with the values for P1 within errors.
This result is naturally understood 
if we assume that P1 and P2 are interface between Al and N plasmas
(see \S~\ref{sec:discussion}). 
Then, the ram pressure of Al plasma
is so large that the presence or absence of the magnetic pressure in the N plasma is negligible
at least in early epoch.
The intensity change is sharper for $B_0=3.6$~T case than $B_0=0$ case;
however, the physical interpretation of the observed width of the intensity gradients
is difficult because there are many possible explanations (see \S~\ref{app:deltax} for details).

For the structure R2,
when we determine the values of $v_0$ and $t_0$,
the situation is similar to the case of R1 in the case of $B_0=3.6$~T case than $B_0=0$.
SOP count variation is small and scale lengths of the intensity gradient of R2 is $\Delta x\approx0.5$~mm,
making it difficult to define the position of R2.
Nevertheless, one can 
 identify the boundary of green region in Fig.~\ref{fig:sop}(c) for $t\approx15$--25~ns.
Assuming again the functional form given by Eq.~(1) with 
$v_0=1000$~km~s$^{-1}$ and $t_0=64$~ns,
we draw the dashed line R2 in Fig.~\ref{fig:sop}(c).
 As in the case of R1, one can see from Fig.~\ref{fig:sop_TE_SE}(c) that 
 around R2, the rate of intensity increase becomes higher than before.
Figure~\ref{fig:selfemi23nsB} shows  the spatial profiles of the self-emission at 
$t=22$ and 23~ns,
 that is, one-dimensional slices of Fig.~\ref{fig:sop}(c).
Despite  fluctuation via instrumental pixel damage and statistical noise, 
we identify an intensity decrease  from $X'=1.45$ to 1.5~cm
at $t=23$~ns at which we performed TS measurement.
Comparing with the profile at $t=22$~ns (black-dashed line),
one can see that R2 propagates outward.

Again as seen in Fig.~\ref{fig:sop}(c), self-emission intensity is inhomogeneous upstream of R2.
This fact is similar to the case of $P_N= 5$~Torr and $B_0=0$.
(see \S~\ref{app:SOPinhomo} for detailed discussion).

\subsection{Analysis of Ion term of Collective Thomson scattering (TS) spectra}

We fit TS spectra assuming the resonance with ion-acoustic waves
(IAWs) in plasmas, which have a single ion-component 
in Maxwelian distribution.
The IAW spectrum is fitted with a convoluted spectral density function
\begin{eqnarray}
	\tilde{S}(\mathbf{k},\omega) &=& \int S(\mathbf{k}, \omega') R(\omega'-\omega) d\omega',\\
	S(\mathbf{k},\omega) &=& \frac{2\pi Z}{k}\left|\frac{\chi_e}{\epsilon}\right|^2 f_i(\frac{\omega}{k}),
\end{eqnarray}
where 
$R$ is the resolution of the spectrometer evaluated from the
Rayleigh scattering and expressed as a Gaussian,
$\chi_e$ and $\epsilon$ are electron susceptibility and longitudinal
dielectric function, respectively,
$\mathbf{k} = \mathbf{k}_s - \mathbf{k}_i$, 
$\omega = \omega_s - \omega_i$,
$f_i$ is the ion distribution function, and
the charge state $Z$ is self-consistently derived 
from a collisional radiative model with the FLYCHK code \cite{Chung2005}.
The TS scattered light intensity is proportional to the electron density $n_e$, and it is determined 
by an absolute calibration of the collective TS system 
using the following formula:
\begin{eqnarray}
	I_T/I_R = \frac{n_e \sigma_T E_T}{n_N \sigma_R E_R}\frac{S_i}{2\pi},
\end{eqnarray}
where $I$, $\sigma$, and $E$ are the light intensity, cross section,
and incident laser energy, respectively,
the subscripts  ``T''  and ``R''   represent the Thomson and Rayleigh scattering,
respectively,
$n_N$ is the nitrogen density for Rayleigh scattering measurement,
and $S_i$ is the total intensity of the ion component
expressed as $S_i = \int S(\mathbf{k},\omega)d\omega$.
Errors for plasma parameters by TS measurements are evaluated 
from best-fitted values and covariance that we get 
from the least-squares fitting of the observed IAW spectra.

Results are summarized in Table~\ref{tbl:LaserParameters}, where we assume N plasma except for 
the case of rapidly moving ($V_{X'}\approx737$~km~s$^{-1}$) component that is seen by IAW-1 at around $\lambda\approx530.2$~nm
at $t=10$~ns, $p=0$ (TCC), and $B_0=0$ (Fig.~\ref{fig:LTS_B=0}(b))
and the case of $t=30$~ns, $p=0$ (TCC), and $B_0=0$ (Fig.~\ref{fig:LTS_B=0}(d)).
For these cases, we show in the table both results on the assumption of N and Al plasmas.

As stated above, we analyzed IAW spectra on the assumption of a single component plasma with Maxwell distribution. 
In our case, the distribution function deviates from Maxwellian around the shock transition layer. 
However, at present, analysis method in such a case has not been established, because we have 
no matured theory of TS spectrum in that case. 
This issue may be a future problem to be resolved in the community of shock experiments. 
Hence, we cannot help but say that derived parameters, listed in Table~\ref{tbl:LaserParameters},  
are just approximate values guiding our theoretical interpretation.


\subsubsection{Case of $P_N= 5$~Torr and $B_0=0$}

\begin{figure}[t]
\centering
\includegraphics[width=100mm, bb= 0 0 950 730]{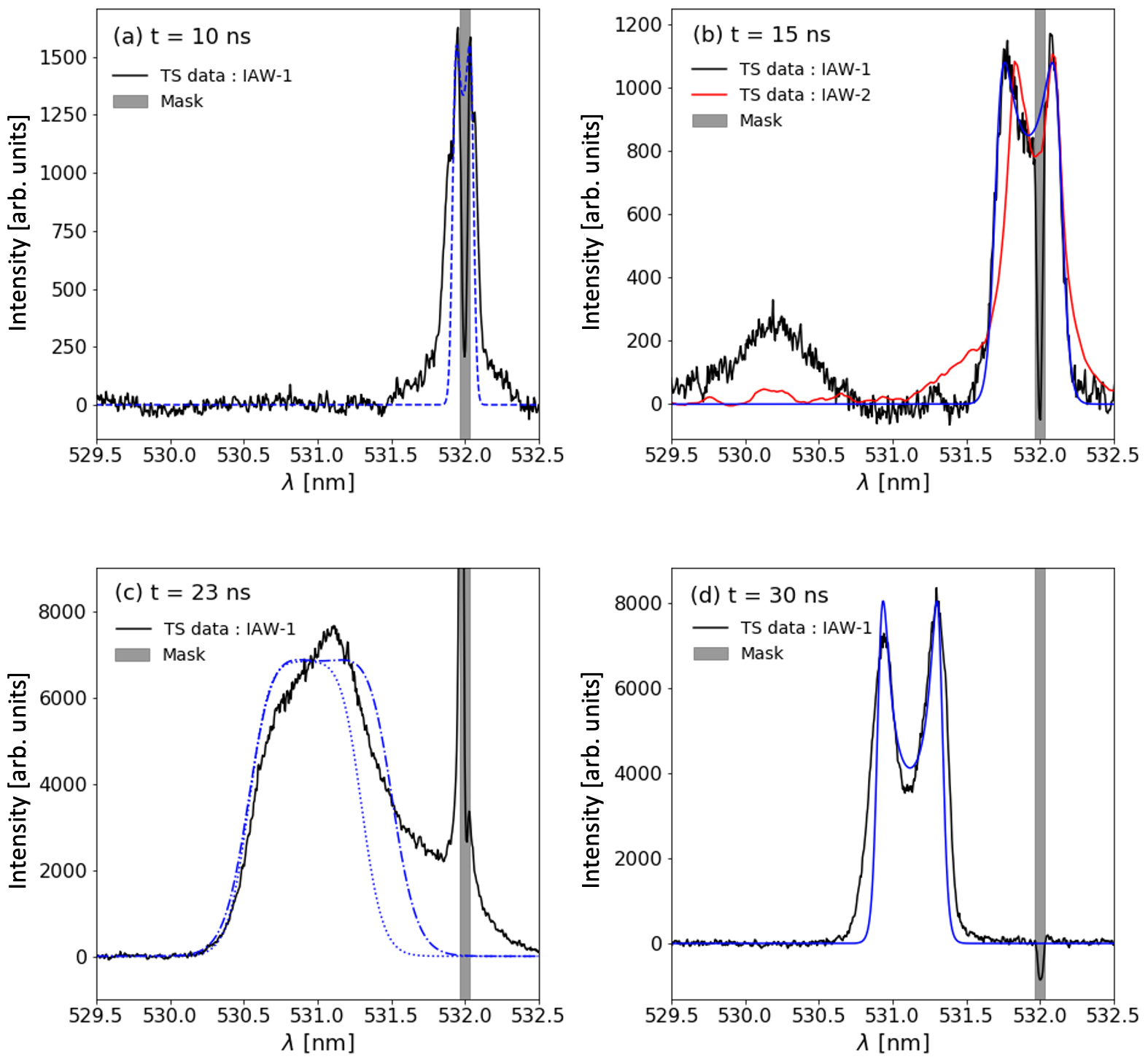}
\caption{
Background-subtracted TS spectra at TCC ($p=0$) obtained at
(a) $t=10$~ns (b) 15~ns, (c) 23~ns, and (d) 30~ns, for the case of $P_N=5$~Torr and $B_0=0$.
Shaded area around the incident wavelength of 532.0~nm is affected by stray light.
Black curves are data of IAW-1, while the red curve in panel (b) is of IAW-2.
Blue solid lines in panels~(b) and (d) show the best-fit results,
while the dashed line in panel~(a) explains only lower-temperature component (see \S~\ref{app:TS10nsB0} for details).
Blue dashed line (dot-dashed line) in panel~(c) is theoretically expected spectrum from N plasma in equilibrium state 
with parameters 
$T_i=1$~keV, $ZT_e=0.73$~keV, $Zn_e=2.6\times10^{19}$~cm$^{-3}$, and $V_{X'}=450$~km~s$^{-1}$
($T_i=1.5$~keV, $ZT_e=1.4$~keV, $Zn_e=4.1\times10^{19}$~cm$^{-3}$, and $V_{X'}=400$~km~s$^{-1}$).
\label{fig:LTS_B=0}
}
\end{figure}

In Figs.~\ref{fig:LTS_B=0}(a) and \ref{fig:LTS_B=0}(b),
two peaks corresponding to the resonance with IAWs are seen  at around 532~nm. 
These are from upstream N plasma almost at rest.
Just before  the arrival of R1 at TCC ($t=10$~ns), we 
identify almost at rest, cold ($T_e$, $T_i<10$~eV) N plasma 
(Fig.~\ref{fig:LTS_B=0}(a): see \S~\ref{app:TS10nsB0} for further discussion).
A few ns after the passage of R1 ($t=15$~ns), 
the static N plasma was heated
($T_e\approx100$~eV and $T_i\approx230$~eV).
We also identify  the moving plasma at $\lambda\approx530.2$~nm from IAW-1 but not from IAW-2,
showing the ion dynamics is collisionless (see also \S~\ref{app:coulombR1}).
Assuming this component is moving along the $X'$-axis, we derive
the bulk velocity $V_{X'}\approx740$~km~s$^{-1}$ (see \S~\ref{app:TS15nsB0}), which is roughly consistent with the velocity of R1 at $t\approx15$~ns.

The TS spectrum taken at $t=23$~ns ($p=0$) does not show a clear double peak (Fig.~\ref{fig:LTS_B=0}(c)).
If we fit the spectrum in the same way as the other epochs, so that assuming
that the observed spectrum is made by IAW in plasma with Maxwell distribution, then we derive
unnatural parameters: namely, electrons would move {\it toward} target with bulk velocity of $\approx1100$~km~s$^{-1}$.
This indicates that the plasma in this epoch and space is nonstationary and/or in highly nonlinear regime.
In any case, the spectrum seems to show high $T_i$ ($\approx 1$~keV: see blue dotted and dot-dashed lines in Fig.~\ref{fig:LTS_B=0}(c)).
The TS spectrum at $t=30$~ns 
reveals plasma with a clear double peak
with $V_{X'}\approx360$~km~s$^{-1}$
(Fig.~\ref{fig:LTS_B=0}(d): see \S~\ref{app:TS30nsB0}).

\subsubsection{Case of $P_N= 5$~Torr and $B_0=3.6$~T}

Next, we analyzed the data of a shot with 
$P_N= 5$~Torr and $B_0=3.6$~T.
Figure~\ref{fig:extmag}(a) shows the TS spectrum at $t=23$~ns,
where the vertical $p$-axis represents the position along the probe laser (see Fig.~\ref{fig:setup2}(c)), so that
$(x,y,z)=(-p\cos45^\circ,p\sin45^\circ,0)$.
A clear double peak of the ion-acoustic resonance is identified when $p>0$.
Assuming N plasma, we fit the spectrum at $p=1$~mm (Fig.~\ref{fig:extmag}(b)) 
and obtain 
$V_{X'}=400$~km~s$^{-1}$ and
the ion density $n_N=n_e/Z=1.2\times10^{18}$~cm$^{-3}$ which is 
about 3.6 times as large as the initial upstream one,
indicating the ion compression. 
Note that the double-peak feature in the TS spectrum still can be seen for $p>1$~mm,
where sensitivity becomes weaker; however, the feature vanishes for $p<0$~mm (Figs.~\ref{fig:extmag}(a) and \ref{fig:TS23nsBwide}).
Such an edge feature corresponds to R2 in Fig.~\ref{fig:sop}(c),
and it is first observed by spatially resolved TS measurement.
In addition, as shown in Table~\ref{tbl:LaserParameters}, 
the density has a maximum at $p\approx1$~mm
(see \S~\ref{app:TS23nsB3.6} for further discussion).


\begin{table*}[b]
\rotatebox{90}{
\begin{minipage}{\textheight}\centering
	\caption{\label{tbl:LaserParameters}%
	Parameters determined by TS measurements (IAW-1) assuming Maxwelian distribution and FLYCHK. 
	We assume N plasma except for the case of fast component with $V_{X'}\approx740$~km~s$^{-1}$ at
	$t=15\pm1.5$~ns and $p=0$ and for the case of $t=15\pm1.5$~ns and $p=0$.
	}
\begin{ruledtabular}
\begin{tabular}{cc|ccccc|ccccc}
&  & 
\multicolumn{5}{c|}{$B_0=0$} & 
\multicolumn{5}{c}{$B_0=3.6$~T} \\
\hline
\hline
$t$\footnote{Time from laser irradiation.} & 
$p$\footnote{$p=0$ corresponds to TCC (see Fig.~\ref{fig:setup2}(c)).} & $T_i$ & 
$T_e$ &
$n_e$ &
$Z$ &
$V_{X'}$ &
$T_i$ & 
$T_e$ &
$n_e$ &
$Z$ &
$V_{X'}$ \\
~[ns]~ & [mm] & [eV] & [eV] & [$\times10^{18}$cm$^{-3}$] & & [km~s$^{-1}$]  &
[eV] & [eV] &  [$\times10^{18}$cm$^{-3}$] & & [km~s$^{-1}$] \\
\hline
$10\pm1.5$ & $-2$ & \multicolumn{2}{c}{$5.9\pm0.4$\footnote{Assuming $T_i=T_e$.}}  & $1.2\pm0.4$ & 2.8 & $4.5\pm0.5$ & \multicolumn{5}{c}{$\cdots$} \\
$10\pm1.5$ & 0\footnote{Cold component.} & $6.9\pm3.1$$^{\rm d}$ & $7.5\pm0.8$$^{\rm d}$ & $1.0$$^{\rm d,f}$ & 3.2$^{\rm d,f}$ & $6.0\pm0.5$$^{\rm d}$ & \multicolumn{5}{c}{$\cdots$} \\
                    &  0\footnote{Warm component.} & $52\pm7$$^{\rm e}$ & $19\pm3$$^{\rm e}$ & $1.6$$^{\rm e,}$\footnote{Fixing N ion density, $n_N=n_e/Z=3.2\times10^{17}$cm$^{-3}$.} & 5.0$^{\rm e,f}$ & $9.9\pm0.7$$^{\rm e}$& \multicolumn{5}{c}{$\cdots$} \\
\hline
$15\pm1.5$ & 0 & $228\pm4$ & $98\pm2$ & $1.7\pm0.4$ & 5.5 & $31\pm0.8$ & \multicolumn{5}{c}{$\cdots$} \\
                    &  0 & $(1.0\pm0.3)\times10^3$ & --- \footnote{Unconstrained.} & $0.22\pm0.05$ & 4.0 & $737\pm2$ & \multicolumn{5}{c}{$\cdots$} \\	
                    &  0\footnote{Assuming Al plasma instead of N plasma.} & $(2.0\pm0.6)\times10^3$\,$^{\rm h}$ & --- $^{\rm g,h}$ & $0.26\pm0.06$$^{\rm h}$  & 4.9$^{\rm h}$ & $737\pm2$$^{\rm h}$ & \multicolumn{5}{c}{$\cdots$} \\
\hline
$23\pm1.5$ & 0 & \multicolumn{5}{c|}{N/A (see text)} & $609\pm23$ &$188\pm6$& $4.9\pm2.3$ & 6.7& $363\pm1.5$ \\
$23\pm1.5$ & 0.5 & \multicolumn{5}{c|}{N/A (see text)} & $222\pm6$ & $171\pm3$ & $5.5\pm1.8$ & 6.5 & $389\pm0.8$ \\
$23\pm1.5$ & 1 & \multicolumn{5}{c|}{N/A (see text)}   & $205\pm6$ & $161\pm3$ & $7.4\pm2.5$ & 6.4 & $394\pm0.8$ \\
$23\pm1.5$ & 1.5 & \multicolumn{5}{c|}{N/A (see text)} & $275\pm8$ & $148\pm2$ & $5.7\pm2.5$ & 6.1 & $386\pm0.7$ \\
\hline
$30\pm1.5$ & 0 & $166\pm5$ & $95\pm2$ &$4.9\pm0.7$ & 5.1  & $363\pm0.6$ & \multicolumn{5}{c}{$\cdots$}\\
                    &  0$^{\rm h}$ & $518\pm8$$^{\rm h}$ & $137\pm2$$^{\rm h}$ & $3.7\pm0.27$$^{\rm h}$  & 10.3$^{\rm h}$ & $364\pm0.4$$^{\rm h}$ & \multicolumn{5}{c}{$\cdots$} 
\end{tabular}
\end{ruledtabular}
\end{minipage}
}
\end{table*}



\section{Discussion}\label{sec:discussion}

We performed 
1D PIC simulations with similar conditions to our experiment.
For  details of the simulation set up, see Umeda et al.~\cite{Umeda2019}.
We adopt the same parameters as those of Run~1 (and Run~2) of Umeda et al.~\cite{Umeda2019}, 
but we consider two cases with different  external magnetic field strength  in
the N plasma, $B_0=0$  and $B_0=3.5$~T.
In the Al plasma, the magnetic field with the strength of 10~T is externally imposed.
This is  a simple artifact of a self-generated Biermann field,
although our  1D simulation  cannot capture the developement of the field.
Note that we use real electron-to-ion mass ratios
$m_i/m_e=49572$ and 25704 for Al and N plasmas, respectively.
Our PIC simulation scheme does not incorporate the effects of Coulomb collisions.

Below we interpret our experimental results comparing with 1D PIC simulations. 
Such an argument is similar to previous experimental studies on supercritical magnetized shocks. 
Note, however, that as already stated in 
\S~\ref{sec:intro},
laser experiments are possibly capable of seeing larger-scale, longer-time evolution of plasma interaction, which is currently unachievable with 3D PIC simulations with real mass ratio. This causes a dilemma when comparing experimental and simulation results since we cannot accurately evaluate multi-dimensional effects (e.g., non-plane-parallel plasma expansion resulting in adiabatic cooling and dilution of the Biermann magnetic field, excitation of obliquely propagating waves in highly inhomogeneous plasmas, and so on). Hence, we should not fully rely on the numerical simulation results, and it is not necessary that experimental results 
perfectly (quantitatively) match the simulation results. Nevertheless, as described below,
our 1D PIC simulation results in both $B_0=0$ and 3.5~T cases are 
broadly consistent with experimental data, so that we believe that our simulations already 
catch essential features related on the magnetized shock formation.

According to our 1D PIC simulation for $B_0=0$ \cite{Umeda2019},
the sharp rise of the electron density $n_e$ occurs at 
an interface between N and Al plasmas
in the electron scale, so that we interpret P1 seen in SOP (Fig.~\ref{fig:sop}(b)) as this electron-scale discontinuity.
This is supported by the fact that high electron and ion temperatures are observed at $t=15$ and 23~ns,
upstream of this interface.
Our radiation hydrodynamics simulation 
(see \S~\ref{app:RHD}) shows that
the Biermann magnetic field in Al plasma
has initial strength $\gtrsim100$~T, and as the plasma expands the field becomes weaker
($\approx10$~T at $t=8$~ns around the head of the Al plasma).
It is at least partially responsible for the N ion reflection  \cite{Umeda2019}.
Collisional coupling is also non-negligible (see \S~\ref{app:coulombP1P2} for details).
Later the interface P1  decelerates due to the interaction between Al and N plasmas, so that
its velocity decreases to $\sim430$~km~s$^{-1}$
after propagating $\sim1$~cm from target (at $t\approx20$~ns).

Our 1D PIC simulation for $B_0=0$
also showed that some Al ions penetrate beyond the interface,
being accelerated by the ponderomotive force \cite{Umeda2019}.
These fast Al ions might correspond to R1 seen in Fig.~\ref{fig:sop}(b) 
(see \S~\ref{app:PIC_B0}  for more discussion).
The fact that the initial velocity of P1 
($v_0\approx590$~km~s$^{-1}$)
is smaller than $v_{\rm Al,0}$
implies that at least some Al ions penetrate upstream beyond P1.
Another possible explanation of R1 is N ions reflected
around the head of Al plasma.
This might be indicated by our experimental result that
 the initial velocity of R1 ($\approx1600$~km~s$^{-1}$)
 is just as twice as the initial Al velocity ($v_{\rm Al,0}\approx800$~km~s$^{-1}$).
During the propagation, such fast ions decelerate due to the interaction with 
upstream N plasma at rest, leading to the heating of incident N plasma between P1 and R1.

\begin{figure}[t]
\centering
\includegraphics[width=100mm, bb=0 0 1040 870]{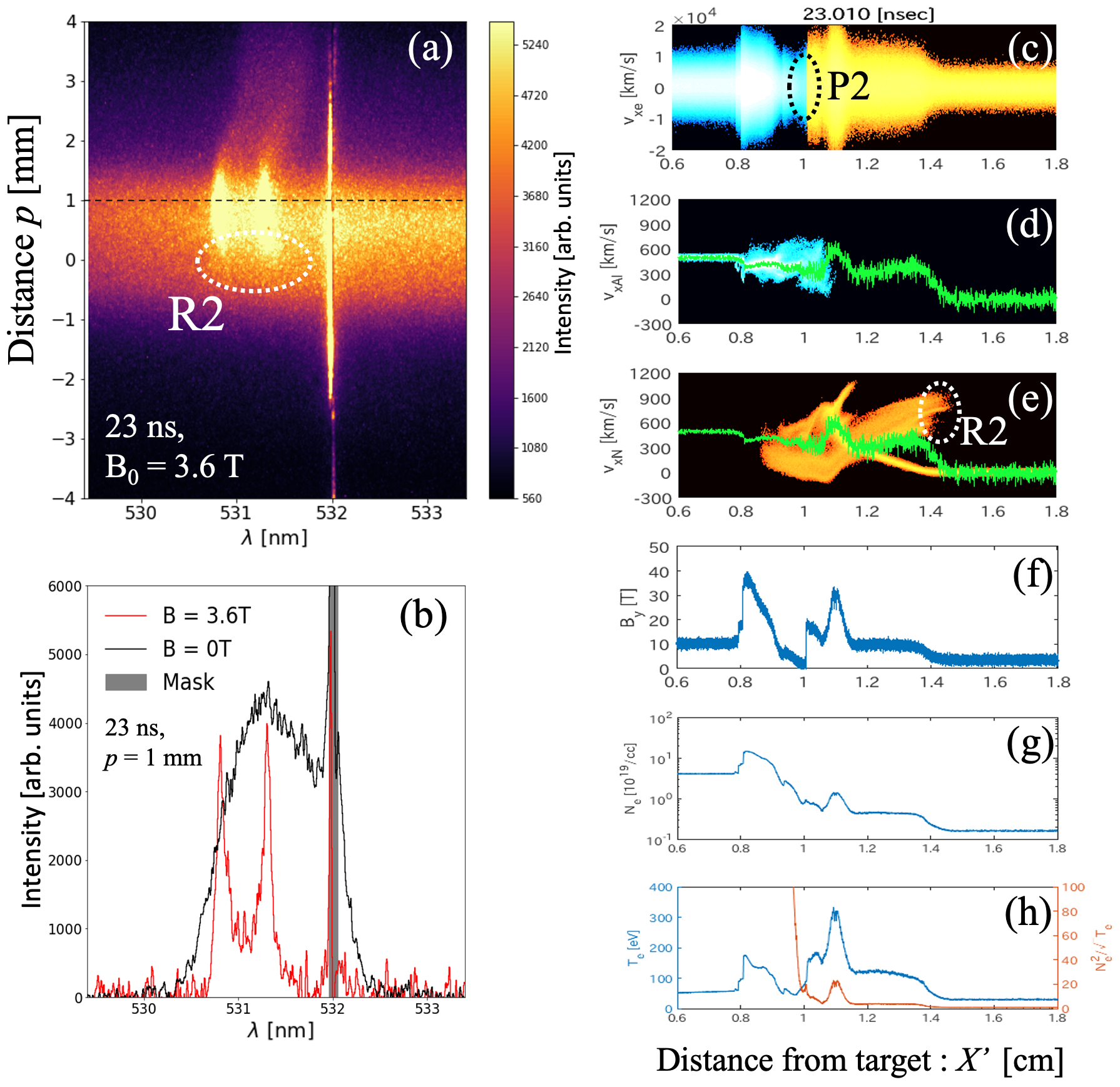}
\caption{
(a) Measured TS spectrum at $t=23$~ns for $B_0=3.6$~T
along the probe laser axis $p$, which was obtained by IAW-1.
(b) Comparison of TS spectra at $p=1$~mm (dashed line in panel~(a)) for $B_0=3.6$~T (red) and 0~T (black) cases.
(c-h)~Results of 1D PIC simulation with $B_0=3.5$~T at $t=23$~ns, in which the
horizontal axis is the distance from the target $X'$.
(c) Electron phase-space plot. 
(d,e) Ion phase-space plots.
In these panels, blue and yellow points represent Al and N plasmas, respectively, and
green curves show the electron bulk velocity.
Bottom three panels show spatial profiles of 
transverse magnetic field strength (f), 
electron density $n_e$ (g), 
electron temperature $T_e$ (blue line in panel~(h)),
and emissivity of the plasma self-emission $n_e^2/\sqrt{T_e}$ normalized by far upstream value (red line in panel~(h)).
\label{fig:extmag}
}
\end{figure}

Our analysis results of SOP and TS measurements for $B_0=3.6$~T case
are again broadly consistent with results of 1D PIC simulation
with $B_0=3.5$~T (Figs.~\ref{fig:extmag}(c)-(h) and Fig.~\ref{fig:PIC_3.5T}).
Observed steep emission gradient P2 in Fig.~\ref{fig:sop}(c) is associated with an interface between Al and N plasmas 
(see Fig.~\ref{fig:extmag}(c) and red line in Fig.~\ref{fig:extmag}(h)).
Incident N ions do not penetrate deeply into the Al plasma because of 
compressed Biermann magnetic field (Fig.~\ref{fig:extmag}(f)) and/or collisional coupling.
On the other hand, Al ions are trapped by compressed external magnetic field, and do not enter the N plasma (Fig.~\ref{fig:extmag}(d)).
Incident N ions are initially reflected at the interface between Al and N plasmas (which is observed as P2 in Fig.~\ref{fig:sop}(c))
and later reflected  by the compressed external field in the N plasma (Fig.~\ref{fig:PIC_3.5T}).
They are gyrating back due to the external magnetic field (Fig.~\ref{fig:extmag}(e)),
forming a shock foot.
Hence, it is natural to interpret that R2 in Fig.~\ref{fig:sop}(c)
corresponds to the edge of the reflected N ions (Fig.~\ref{fig:extmag}(e)),
which has also been confirmed by TS measurement (Fig.~\ref{fig:extmag}(a)).
The density $n_e$ takes maximum at $ p\approx1$~mm (Table~\ref{tbl:LaserParameters}), which is about to be a  shock overshoot.
Observed $T_i\approx0.2$~keV at $p\approx1$~mm
of the reflected component is larger than incident N ions,
but is smaller than that for $B_0=0$ (Fig.~\ref{fig:extmag}(b)).
This is also consistent with PIC results.
Therefore, we claim that the spatially resolved edge of the foot of a 
developing magnetized collisionless shock is measured
(see \S~\ref{app:TS23nsB3.6}, \S~\ref{app:PIC_B3.6}, and \S~\ref{app:coulombR2} for further discussion).

We estimate upstream physical quantities.
We had only one shot with $B_0=3.6$~T, and TS spectrum at $t=23$~ns.
Hence, we simply assume the upstream plasma parameters except for $B_0$ are the same as those for the
unmagnetized case.
We analyze the data of  IAW-1 for the case of $P_N=5$~Torr and $B_0=0$
at $t=10$~ns and $p=-2$~mm, at which the discontinuity R1 had not arrived yet
(white triangle in Fig.~\ref{fig:sop}(b)).
We fit the TS spectrum to get
the N ion density $n_N=n_e/Z=(4.4\pm1.5)\times10^{17}$~cm$^{-3}$ (see \S~\ref{app:TS10nsB0}), 
suggesting the upstream medium is fully ionized.
Using the best-fitted values, we get upstream quantities such as
the sound speed $a_{S}\approx 11$~km~s$^{-1}$, 
Alfv\'{e}n velocity $v_{A} \approx 32$~km~s$^{-1}$, 
and ion (electron) plasma beta $\beta_i\approx0.08$ ($\beta_e\approx0.2$) for $B_0=3.6$~T.
Simply assuming magnetohydrodynamics,
we expect that the velocity of a forming shock
may be higher than that
 of interface between Al and N plasmas (P2: see Figs.~\ref{fig:sop}(c) and \ref{fig:extmag}(c)), 
whose typical value is $v_{ej} \approx400$~km~s$^{-1}$. 
Then, we expect the magnetosonic Mach number
$M_{ms} > v_{ej}/\sqrt{v_A^2+a_S^2} \approx 12$ and
Alfv\'{e}n Mach number
$M_A> v_{ej}/v_A \approx 13$, so that the shock will be
 supercritical with ion-ion mean-free path
$\lambda_{ii} = m_i^2 v_{ej}^4/8\pi n_N Z^4 e^4 {\rm ln}\Lambda \approx 3.5$~cm.

\section{Summary}

We have conducted experiments of generating collisionless
shocks propagating into
magnetized plasma at rest with uniform mass density.
It is quite difficult to identify a shock formation only from SOP
due to the bright emission enhancement at P2.
However, our PIC simulations show that R2 is composed only of N plasma, and
if we analyze the TS data assuming the N plasma, all the experimental data
self-consistently show that R2 is an edge of the foot of a forming supercritical shock in the magnetized N plasma.
We are also sure that a shock ramp and overshoot are arising  between P2 and R2 as seen
in the PIC result.



\appendix
\section{Notes on SOP Analysis}

\subsection{On the inhomogeneity upstream of R1 and R2}\label{app:SOPinhomo}

As shown in Figs.~\ref{fig:sop}(b) and \ref{fig:sop}(c), self-emission intensity is not uniform both upstream of R1 in 
$B_0=0$ case and of R2 in $B_0=3.6$~T case. This indicates the ionization of upstream N plasma is not uniform. 
Ahead of R1 and R2, neither Al nor reflected N plasma co-exist, so that the cause of the non-uniformity is likely time-dependent photoionization. 
Around the interface between Al and N plasmas (P1 and P2), the electron temperature is high and it can be more than 100 eV as shown by our
1D PIC simulations (Figs.~\ref{fig:pic23ns0T} and \ref{fig:PIC_3.5T}).
Such high-density, hot plasma emits ionizing photons, and they change the ionization state of the upstream N plasma. 
When photoionized, N ions eject hot electrons, resulting in the plasma heating. 
This causes further change of the upstream ionization state.
Both collisionless and collisional processes depend on the ion charge state.
However, the upstream number density of N ions 
(i.e., the upstream mass density, which determine the whole shock dynamics from ion scale to hydrodynamical scale) is still uniform.


\subsection{Observed width of the density gradients}\label{app:deltax}

In ideal plane-parallel case, width of the transition layers P1 and P2, which are the interfaces 
between the Al and N plasmas in electron scale, is given by the gyro radius of thermal electrons around the layers
if Al and N interaction is collisionless.
It is estimated as $r_{g,e} = m_e c v/eB\sim3\times10^{-4}{\rm mm}\,(B/10~{\rm T})^{-1}(v/500~{\rm km}~{\rm s}^{-1})$.
The mean-free path of electrons via electron-electron Coulomb collision is on the same order or less.
These lengths are much smaller than the observed apparent width $\Delta x\sim0.3$~mm or slightly less. 
This discrepancy is explained by the three-dimensional (3D) effect: 
geometry of the plasma self-emission  in 3D space is not plane parallel, and even the interface fluctuates during the propagation. 
 In such an actual case, the emission is projected onto the detector plane whose normal vector is in the $y$ direction. 
 As an extreme toy case, let us consider spherically symmetric uniformly emitting sphere with radius $\ell_0$ 
 whose emissivity is given by step function, $f(r) = const$ for $r < \ell_0$ and $f(r) =0$ for $r > \ell_0$ 
 ($r$ is the radial coordinate in the spherical coordinate system). 
 When such emission is projected onto the detector plane, the apparent spatial profile does not remain step-function-like, 
 but has gradual transition whose width is on the order of the curvature radius $\ell_0$. 
 More generally, in cylindrically symmetric case, the projected emission profile is related to the emissivity in the three-dimensional space by Abel transform. 
 In the present case, the exact plane parallel assumption is clearly not a good approximation, 
 although the local curvature radius of the interface is highly uncertain. 
 However, the length on the order of 0.1~mm is not unnatural. 
 These projection effect may become important for not only  P1 and P2 but also the other emission gradients R1, R2 and P0.

There are several  other possibilities to explain the observed width $\Delta x$ of P1 and P2, such as Rayleigh-Taylor instability
and the inhomogeneous expansion of the Al plasma (that is, anisotropic velocity distribution of kinetic energy of Al plasma).
As another case, when the Biermann field in Al plasma is weak ($\lesssim1$~T), the diffusion length of electrons in the direction perpendicular to
the Biermann field is potentially comparable to the observed scale width $\Delta x$.
At present, precise description of the observed width is difficult since there are several physical possibilities.


\section{Notes on Analysis of TS spectra}

\subsection{TS spectra at $t=10$ ns for $B_0=0$}\label{app:TS10nsB0}

\begin{figure}[ht]
\centering
\includegraphics[width=90mm, bb= 0 0 1100 700]{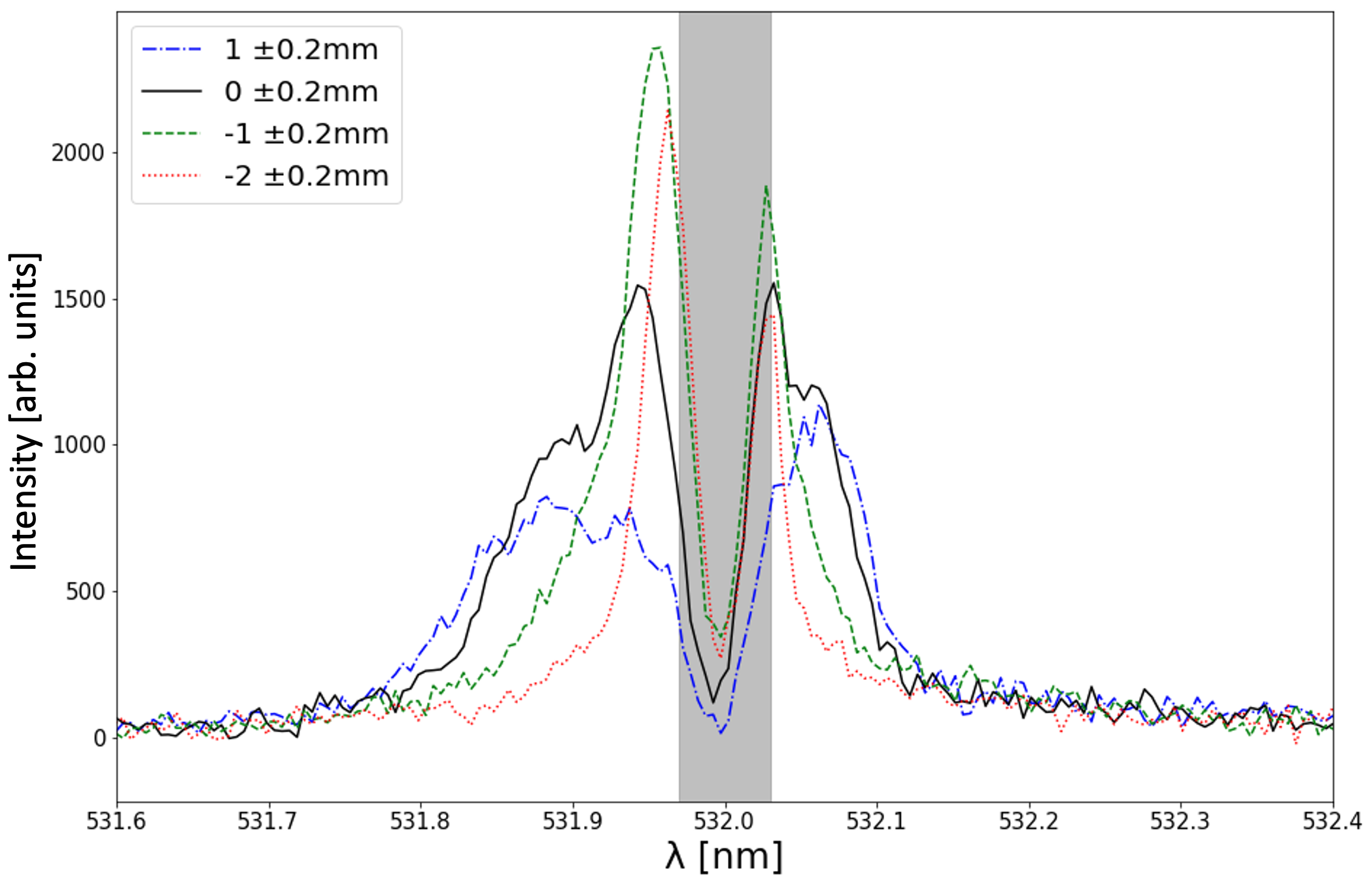}
\caption{
Background-subtracted 
IAW-1 spectra at $t=10$~ns in the case of $P_N=5$~Torr and $B_0=0$ for positions $p=-2$~mm (red dotted line), $-1$~mm (green dashed line), 
0 (black solid line), and 1~mm (blue dot-dashed line).
The intensity in the gray-shaded region around the incident wavelength of 532.0~nm is 
diminished by a filter in the spectroscopic optics system to cut the stray light. TCC corresponds to $p=0$, so that
the black solid line is identical to the data of Fig.~\ref{fig:LTS_B=0}(a).
\label{fig:IAW10ns}
}
\end{figure}

We show in Fig.~\ref{fig:IAW10ns}  the IAW-1 (see Fig.~\ref{fig:setup2}(c)) spectra 
at $t=10$~ns around $\lambda=532$~nm with different positions $p=-2$, $-1$, 0, and 1~mm.
The black solid line  in Fig.~\ref{fig:IAW10ns} ($p=0$, TCC) is identical to the black solid line in Fig.~\ref{fig:sop}(a).

We see TS spectra from N plasma at rest. During the instrumental gate width of 3~ns, the rapid plasma component R1 (shown in Fig.~\ref{fig:sop}(b)) 
passed around TCC. Such rapidly moving plasma has a large Doppler shift, so that its scattered light does not exist around $\lambda\approx532$~nm, 
and may be even outside of the whole range of wavelength coverage of our spectrometer system. 
(If we assume the plasma velocity corresponding to R1 as $V_{X'}=dX'/dt\approx933$~km~s$^{-1}$ with $v_0=1600$~km~s$^{-1}$ and $t_0=48$~ns,
then the Doppler shift $\Delta\lambda\approx 2.3$~nm is predicted.)
The component R1 electrostatically interacts with the incident N plasma, resulting in rapid heating of the N plasma. 
This is also suggested by the rapid increase of self-emission at TCC (see Fig.~\ref{fig:sop_TE_SE}(b)).

The position $p=-2$~mm was in the region at which the rapid component R1 had not arrived yet, 
so we interpret that the TS spectrum at $p=-2$~mm comes from the upstream N plasma that was unaffected by Al ejection. 
This is also justified by our 1D PIC simulation (see \S~\ref{app:PIC_B0}).
Then we fitted the spectrum assuming $T_e=T_i$, and best fitted parameters are shown in Table~\ref{tbl:LaserParameters}.
At $p=-1$~mm, we also see such a cold component, but slightly heated compared with the case of $p=-2$~mm.
At $p=1$~mm, which becomes the downstream region of R1 near the end of exposure of TS measurement, 
one can see warmer plasma with broader separation between double peaks of IAW resonance.

The TS spectrum at $p=0$ (TCC) looks complicated.
It seems to consist of the two superimposed components (cold and warm N plasmas).
First, we fit the spectrum in the wavelength range $531.92~{\rm nm}<\lambda<532.05$~nm
(excluding the gray-shaded region shown in Fig.~\ref{fig:IAW10ns} that is affected by a filter to cut the stray light)
in order to explain only the cold component with narrower separation of the double peak.
Assuming $T_i=T_e$ and $V_{X'}=0$, we derived 
$T_i=T_e=9.3\pm0.2$~eV,
$n_e=(7.3\pm1.6)\times10^{17}$cm$^{-3}$, and $Z=4.2$,
which was represented by the blue dashed line in Fig.~\ref{fig:LTS_B=0}(a).
Then one can clearly see that the fit outside of the fitting wavelength range is inadequate, which indicates the existence of
another warm component.
Hence, we next fit the TS spectrum with two components that are simply superimposed, where we assume
the two components are recorded in the instrumental gate width of 3~ns.
In fitting the data, we fix the N ion density $n_N=n_e/Z=3.2\times10^{17}$cm$^{-3}$.
The best-fitted parameters are shown in Table~\ref{tbl:LaserParameters}.
Note that both cold and warm components are almost at rest, so that both are highly likely N plasmas.

\subsection{TS spectrum at $t=15$ ns for $B_0=0$}\label{app:TS15nsB0}

When we analysed rapidly moving component for $t=15$~ns, $p=0$ and $B_0=0$ 
that was seen in the IAW-1 spectrum (Fig.~\ref{fig:LTS_B=0}(c))
and has a peak at $\lambda\approx530.2$~nm,
we assumed two cases of  Al and N plasmas.
Table~\ref{tbl:LaserParameters} shows the results. In either cases, the plasma has a bulk velocity of
$V_{X'}\approx740$~km~s$^{-1}$, and the ion temperature is on the order of keV.


\subsection{TS spectrum at $t=30$ ns for $B_0=0$}\label{app:TS30nsB0}

Figure~\ref{fig:sop}(b) shows that at $t\approx30$~ns, the interface P1 passes
through TCC ($X'=1.4$~cm, $p=0$).
Hence, the observed TS spectrum may either come from Al or N plasma, so that we fit the spectrum for both cases.
Indeed, it is hard from the best-fitted values shown in Table~\ref{tbl:LaserParameters}
to judge which plasma is responsible for the observed spectrum.
In the case of N plasma, both the electron density $n_e\approx5\times10^{18}$cm$^{-3}$ and the ion density 
$n_N=n_e/Z\approx1\times10^{18}$cm$^{-3}$ are several times as large as initial N plasma densities.
This fact is naturally explained by 1D PIC simulation.
However, we cannot exclude the cases of Al plasma.


\subsection{TS spectrum at $t=23$ ns for $B_0=3.6$~T}\label{app:TS23nsB3.6}

\begin{figure}[ht]
\centering
\includegraphics[width=100mm, bb= 0 0 900 870]{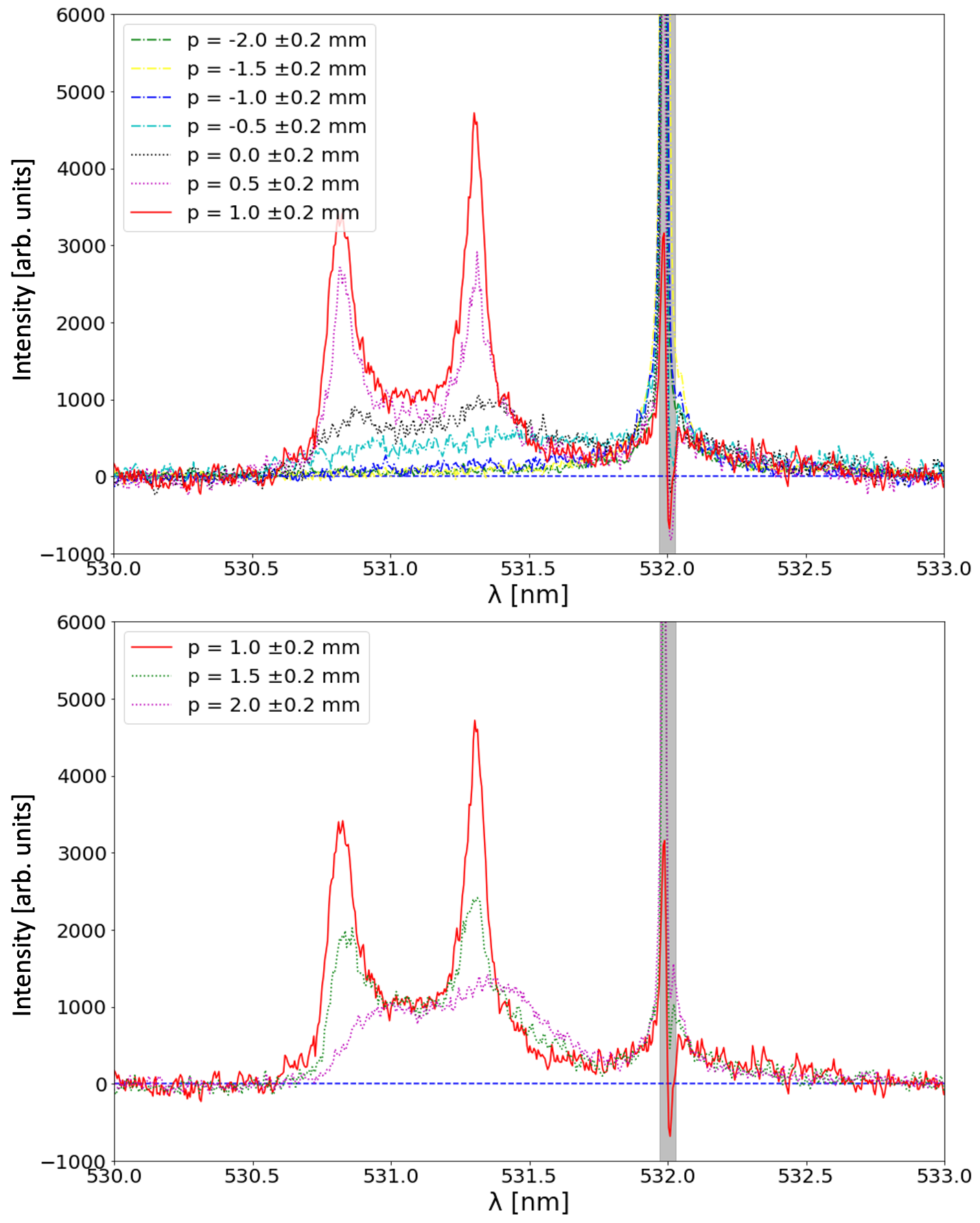}
\caption{
Background-subtracted IAW-1 spectra at $t=23$~ns in the case of $P_N=5$~Torr and $B_0=3.6$~T for various positions $p$.
The upper panel is for cases of $p=-2.0$, $-1.5$, $-1.0$, $-0.5$, 0.0, 0.5, and 1.0~mm, 
while  the lower panel for cases of $p=1.0$, 1.5, and 2.0~mm.
TCC corresponds to $p=0$.
The gray-shaded region around the incident wavelength of 532.0~nm is the same as
that of Fig.~\ref{fig:IAW10ns}.
\label{fig:TS23nsBwide}
}
\end{figure}

\begin{figure}[ht]
\centering
\includegraphics[width=120mm, bb= 0 0 900 360]{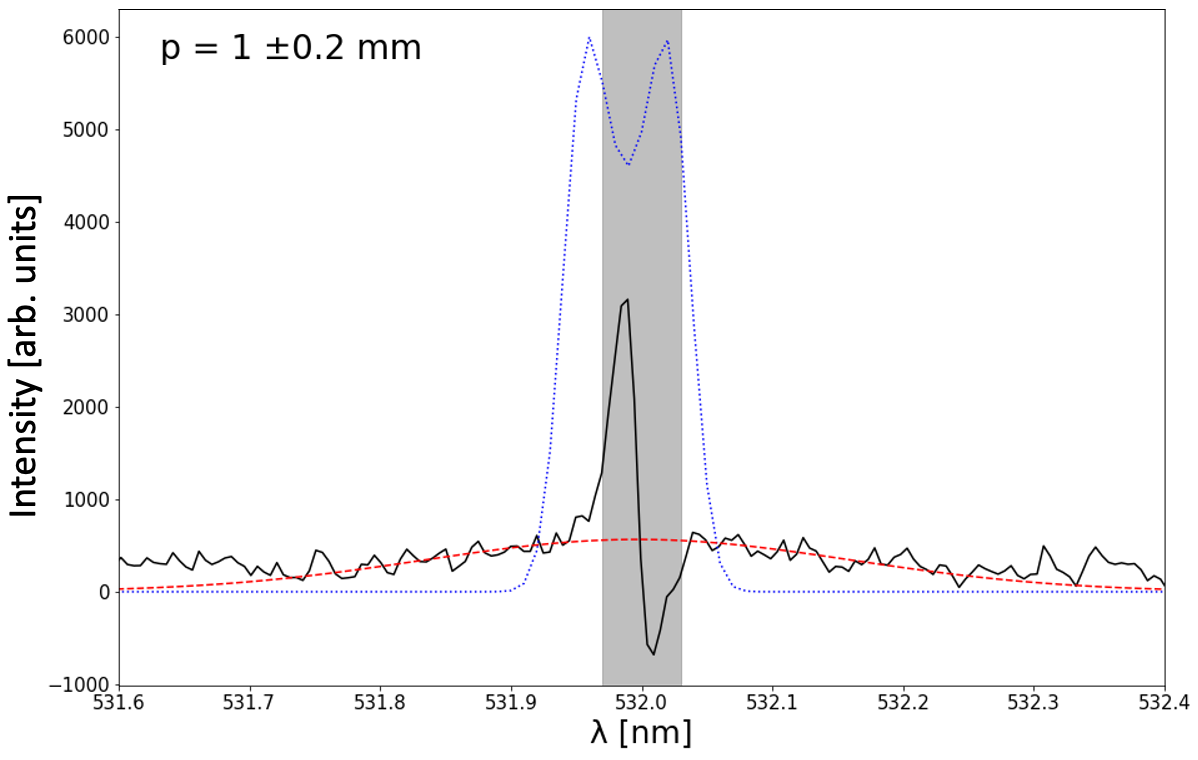}
\caption{
Enlarged view of Fig.~\ref{fig:TS23nsBwide} around 532~nm for $p=1$~mm.
The solid black line is background-subtracted observed data, which is the 
same as red-solid lines in Fig.~\ref{fig:TS23nsBwide}.
The gray-shaded region around the incident wavelength of 532.0~nm is the same as
those of Figs.~\ref{fig:IAW10ns} and \ref{fig:TS23nsBwide}.
The red-dashed line in this figure represents the best-fitted result (see text for details).
The blue-dotted line is expected in the case of the same parameters
as $t=10$~ns, $P_N=5$~Torr, $B_0=0$, and $p=-2$~mm
(i.e., $T_i=T_e=5.9$~eV, $n_e=1.2\times10^{18}$cm$^{-3}$, and $Z=2.8$).
\label{fig:TS23nsB532}
}
\end{figure}

If R2 is the edge of reflected N ions, there should be another N ion population at rest as seen in 1D PIC simulation 
at around $\lambda=532$~nm in the TS spectrum  (e.g., \cite{Schaeffer2019}).
Since the observed data contains bright self-emission as a background and there is intense stray light around the probe laser wavelength ($\lambda=532$~nm), 
it is difficult to perform the spectral analysis to find such a component. 
Using data in the wavelength ranges $\lambda=529.5$--530~nm and 533--534~nm, we determined the background by fitting with a cubic function.
The background-subtracted spectra are shown in Fig.~\ref{fig:TS23nsBwide}. 
One can find small excess around 532~nm. 
Assuming the TS ion feature from N plasma,
we fitted the excess component. 
For $p=1$~mm,
we obtained $Zn_e\approx3\times10^{17}$~cm$^{-3}$ and $T_i\gtrsim1\times10^2$~eV, though
the electron temperature is unconstrained since the observed intensity is weak
 (red-dashed line in Fig.~\ref{fig:TS23nsB532}). 
 Then, we calculate a parameter
 $\alpha = 1/k\lambda_{D,e}\approx0.25(Z/3)^{-1/2}(T_e/100~{\rm eV})^{-1/2}$, 
 where $\lambda_{D,e}$ is the electron Debye length, 
and hence the ion-term scattered light becomes dim.
We also estimate a parameter $\beta = (ZT_e/T_i)^{1/2}[\alpha^2/(1+\alpha^2)]^{1/2}$
that determines the shape of the ion term \cite{sheffield2010plasma}, and we have
$\beta\approx\alpha(ZT_e/T_i)^{1/2}\lesssim0.4(T_i/100~{\rm eV})^{-1/2}$ for small $\alpha$, 
so that it is less than unity as long as $T_i\gtrsim100$~eV.
In such cases, the TS ion term has a single peak as shown by the red-dashed line in Fig.~\ref{fig:TS23nsB532}.

In the case of the fitting result shown above (red-dashed line in Fig.~\ref{fig:TS23nsB532}),
the derived value of $Zn_e$ is about one order of magnitude smaller than expected. 
 If there were the same N plasma with $Zn_e\approx3\times10^{18}$~cm$^{-3}$
 as seen for the case of $B_0=0$, $t=10$~ns, and $p=-2$~mm (see Table~\ref{tbl:LaserParameters}),
 which we refer to as upstream N plasma in this paper, 
 then the TS spectra would have seen by the blue-dotted curve in Fig.~\ref{fig:TS23nsB532}. 
 If we rely on 1D PIC simulation results, the electron temperature alone increases to more than 100~eV around $p=1$~mm.
 In this case, the ion acoustic peaks in the TS spectrum would be more separated from 532~nm,
 so that we would more clearly see the component although the ion-term scattered light would be dim because
the scattering is in the noncollective regime ($\alpha\approx0.8$). 
Hence, we could not experimentally identify the coexistence of incident N plasma almost at rest with low ion temperature and electron density of 
 $\sim10^{18}$cm$^{-3}$.  From the experimental view point,
a probable explanation of  the observed excess around 532~nm is the stray light of the probe laser.

As discussed above, in our present experiment we could not identify another N ion population almost at rest,
coexisting with the R2 component, which is indicated in 1D PIC simulation.
However, we should not directly believe the result of 1D PIC simulations in this case. 
In the 2D or 3D case, plasma interaction proceeds more rapidly, and incident N plasma becomes dilute in velocity space.
As a result, the two components are mixed with each other and merge into a single population in the velocity space.
Another possibility to disentangle this issue is that the plasma temperatures, $T_e$ and $T_i$, of the expected component are
smaller than $\approx6$~eV. Then, TS spectrum from such a cold plasma has narrow peak, and in our present spectrometer system,
it is masked by a filter to cut  stray light.
Such a case potentially occurs because the upstream state is inhomogeneous (see \S~\ref{app:SOPinhomo}) ---
the less ionized are N ions, the less hot electrons, resulting in the lower temperatures and less electron density. 
There may also be the collisional effects.
Finally, we emphasize again that in our analysis of TS ion term, we have assumed Maxwell distributions of ion and electrons and
radiative-collisional equilibrium, which are clearly violated in the shock transition layer.
The derived plasma parameters by our present analysis might be different from actual values.

\section{Estimate of the Biermann Battery Magnetic field}\label{app:RHD}

We performed 2D radiation hydrodynamics simulation \cite{Ohnishi2012}
to have electron pressure and density of Al plasma ejected by laser irradiation.
Then, in order to estimate the strength of the Biermann magnetic field in the Al plasma,
we solve,  as a post process, the induction equation with Biermann term
\begin{eqnarray}
\frac{\partial \bm{B}}{\partial t} &=& \bm{\nabla} \times (\bm{v} \times \bm{B}) + c \frac{\bm{\nabla}p_e \times \bm{\nabla} n_e}{n^2_e e}~~,
\end{eqnarray}
to get the evolution of the field.
It is found that until 4~ns from the shot, the magnetic field strength in the Al plasma
is more than 100~T at the point where the electron density is equal to the critical density of $1.0\times10^{21}$~cm$^{-3}$.
Note that even in this case, the plasma beta is not less than unity, so that 
the generated magnetic field does not alter  the overall dynamics of Al plasma.
After that, the plasma expands and the field strength becomes small (e.g., $\approx10$~T at $t\approx8$~ns
around the head of Al plasma); however, it is
still strong enough to reflect N ions into Al plasma.
Our simulation result on the field generation is quantitatively consistent with previous numerical studies
\cite{Schoeffler2014,Schoeffler2016,Fox2018}.


\section{Details of 1D PIC simulation results}

\subsection{Result for $B_0=0$~T}\label{app:PIC_B0}

This case is identical to Run~2 of  Umeda et al.~\cite{Umeda2019}.
Results for $t\approx14$~ns was presented in Fig.~4 of Umeda et al.~\cite{Umeda2019}, which 
shows the electron-scale tangential discontinuity 
at $X'=0.62$~cm. The electron density rapidly increases there. We interpret this discontinuity as corresponding to
the structure ``P1'' found in SOP.

\begin{figure}[t]
\centering
\includegraphics[width=100mm, bb= 0 0 570 600]{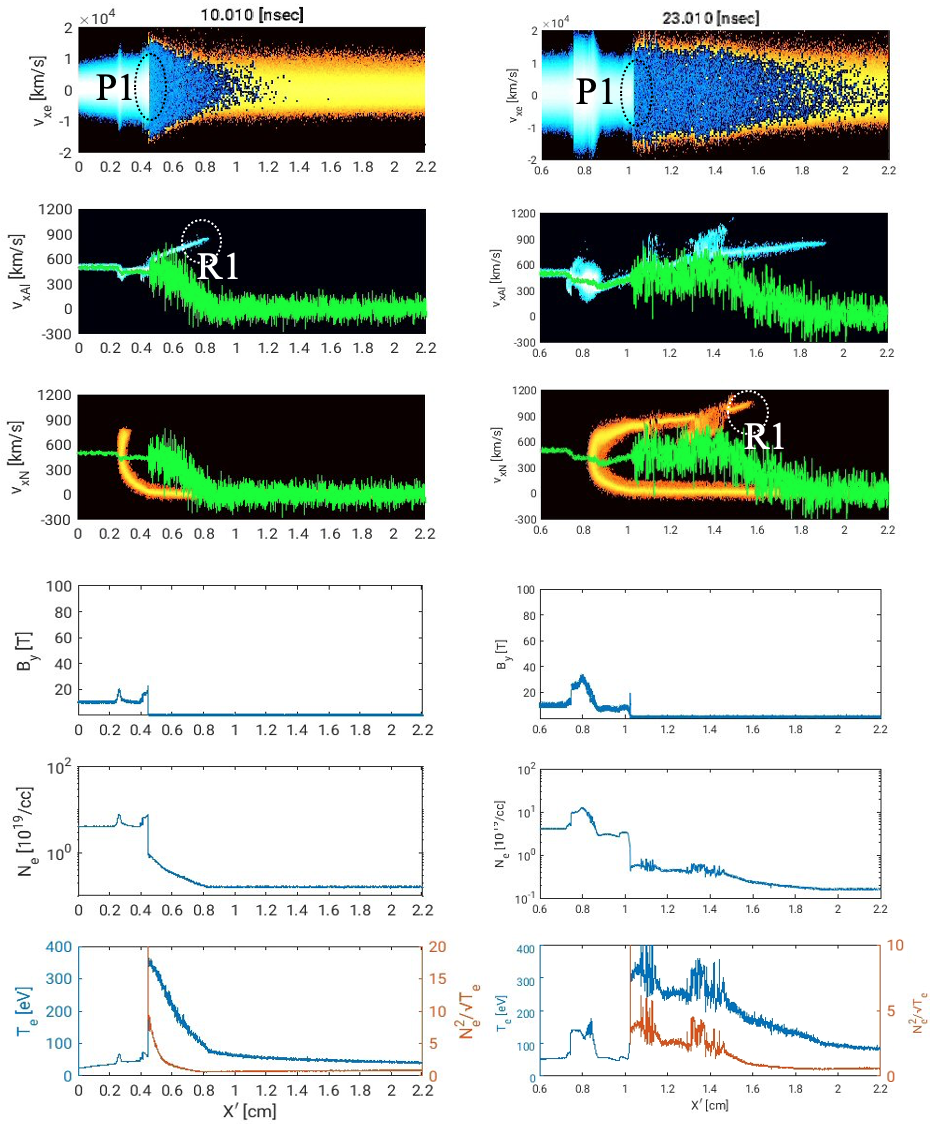}
\caption{
Results of 1D PIC simulation with $B_0=0$~T at $t\approx10$~ns (left) and $t\approx23$~ns (right).
Format is the same as Figs.~\ref{fig:extmag}(c)--\ref{fig:extmag}(h).
The horizontal axis is the distance from the target $X'$.
From top to bottom,
panels of the first row are  electron phase-space plots, and
those of the second and third rows  are ion phase-space plots.
In these panels, blue and yellow points represent Al and N plasmas, respectively, and
green curves show the electron bulk velocity.
Panels of the fourth and fifth rows show the transverse magnetic field strength $B_y$ and
electron density $n_e$, respectively.  
The bottom panels represent electron temperature $T_e$ (blue line)
and emissivity of the plasma self-emission $n_e^2/\sqrt{T_e}$ normalized by far upstream value (red line).
Note that the electron density $n_e$ is displayed with a logarithmic scale.
The bottom panel represents electron temperature $T_e$ (blue line)
and emissivity of the plasma self-emission $n_e^2/\sqrt{T_e}$ normalized by far upstream value (red line).
Note that to enlarge the small variation around a structure ``R1'',
we take the range of the plasma self-emission $n_e^2/\sqrt{T_e}$ up to 20 (left panel) or 10 (right panel), which are different from 
plots in Figs.~\ref{fig:extmag}(h) and  Fig.~\ref{fig:PIC_3.5T}.
\label{fig:pic23ns0T}
}
\end{figure}

In Fig.~\ref{fig:pic23ns0T} we show the results for $t\approx10$~ns and $t\approx23$~ns, the latter of which can be directly compared with
Figs.~\ref{fig:extmag}(c)--\ref{fig:extmag}(h).
A part of Al ions penetrates into incident N plasma after accelerated by ponderomotive force at the interface (P1) between
Al and N plasma. At  $t\approx10$~ns, it reaches  the position $X'\approx0.8$~cm.
As shown in the electron phase space plots (top panels of Fig.~\ref{fig:pic23ns0T}),
a small fraction ($\lesssim1$\%) of electrons originally associated with Al ions, shown by blue dots in the right-side region of P1, that is, 
electrons injected at the left boundary $X'=0$, go across the interface P1.
(In the phase space plots, deeper blue means less number of particles, while light blue or nearly white regions indicate high particle density.)
Nitrogen ions are being reflected by the Biermann magnetic field in Al plasma; however, its edge is still in the Al plasma 
(that is, in the left region of the interface P1 at $X'\approx0.45$~cm).
Interaction between the penetrating Al plasma and incident N plasma is responsible for plasma heating and enhancement of the
plasma self-emission upstream of P1 ($X'\approx0.45$--0.8~cm).
Hence, we interpret the tip of penetrating Al plasma  corresponds to the structure ``R1'' found in SOP in the early epoch.
In our 1D PIC simulation, the region around TCC ($X'=1.4$~cm) is unaffected by the penetrating Al plasma.
Namely, the position $p=-2$~mm (definition of $p$-axis is shown in Fig.~\ref{fig:setup2}(c))
corresponds to $X'\approx1.7$~cm, hence it is expected
that the observed IAW-1 TS data at $t=10$~ns and $p=-2$~mm is appropriate to estimate the parameters of initial upstream N plasma.

At  $t\approx23$~ns, penetrating Al plasma  reaches the position $X'\approx1.9$~cm.
Nitrogen ions are reflected by Biermann magnetic field in Al plasma, and the edge of the reflected component arrives at $X'\approx1.6$~cm.
Before reflected N ions come back, incident N plasma is heated just after the passage of Al ions
by electrostatic interaction ($X'\approx1.6$--1.9~cm), and electron temperature $T_e$ becomes larger.
When we calculate the spatial profile of the intensity of the plasma self-emission (assuming the bremsstrahlung emission, $\propto n_e^2/\sqrt{T_e}$),
we find that it starts to increase at 
the edge of the reflected N ions at $X'\approx1.6$~cm.
Hence, we interpret this edge corresponds to the structure ``R1'' found in SOP in later epoch.

Our 1D PIC simulation used typical parameters of Al and N plasmas 
around TCC ($X'=1.4$~cm) at late epoch (e.g., $t\simeq20$--30~ns) \cite{Umeda2019}. 
The high-speed flow at 1600~km~s$^{-1}$, labeled as  ``R1'',
 is not seen in the 1D PIC simulation. 
On the other hand, parameters of Al plasma and N plasma around the target 
in the experiment are quite different from those of the 1D PIC simulation. 
For example,
Biermann magnetic field in Al plasma is estimated as $\gtrsim100$~T  
at the initial state of the experiment by radiation hydrodynamics
simulation. Such a strong magnetic field may generate a high-speed 
flow at 1600~km~s$^{-1}$ in 2D/3D system.
Then, the position of the emission edge ``R1''  found in SOP
may be much farther from the target.
Nevertheless, it can be said that our PIC simulation qualitatively well explains our experimental results.

\subsection{Result for $B_0=3.5$~T}\label{app:PIC_B3.6}

\begin{figure*}[t]
\centering
\includegraphics[width=430mm, bb= 0 0 2000 850]{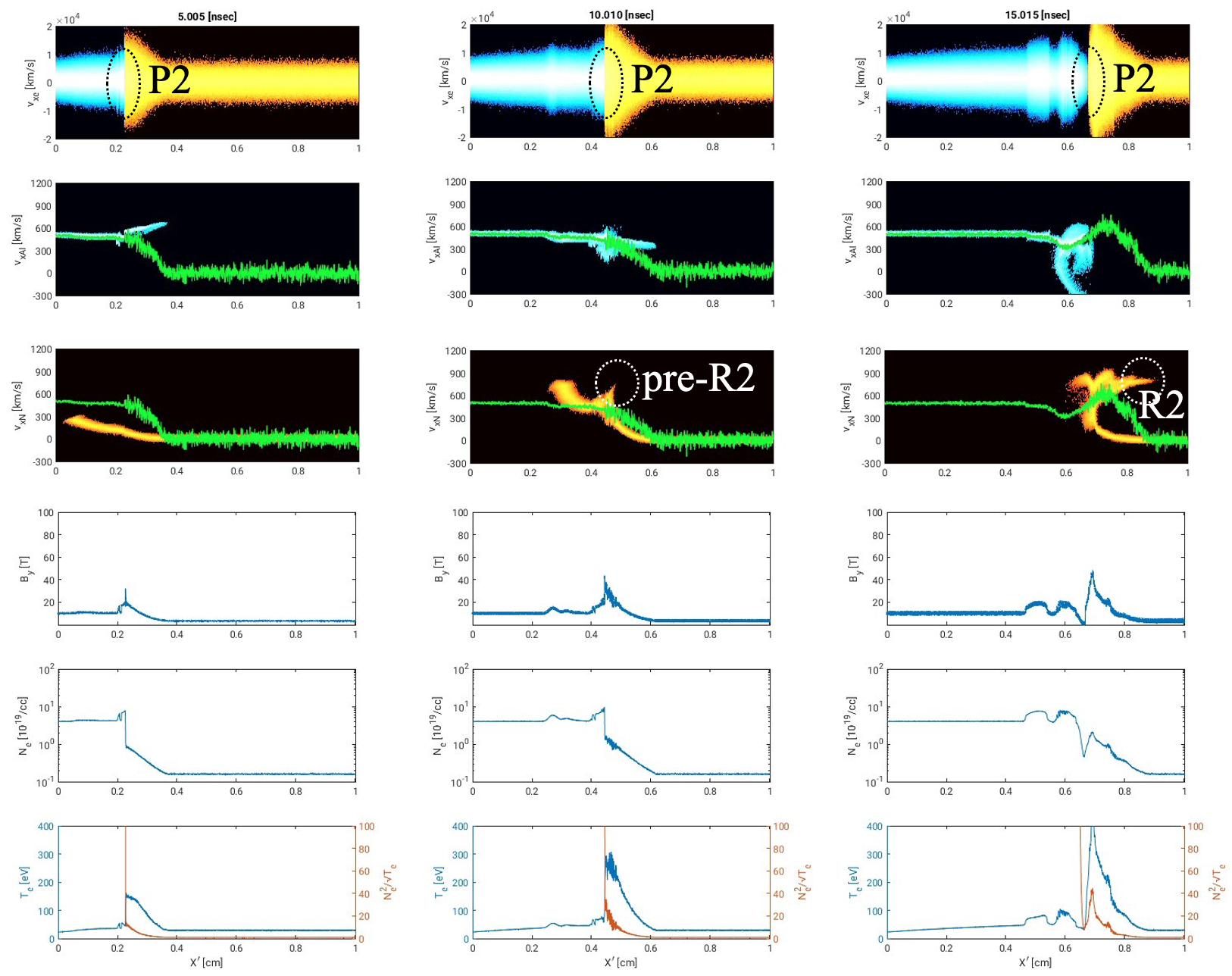}
\caption{
Results of 1D PIC simulation with $B_0=3.5$~T at $t\approx5$~ns (left), 
$t\approx10$~ns (center), and $t\approx15$~ns (right).
Format is the same as Figs.~\ref{fig:extmag}(c)--\ref{fig:extmag}(h).
The horizontal axis is the distance from the target $X'$.
From top to bottom,
panels of the first row are  electron phase-space plots, and
those of the second and third rows  are ion phase-space plots.
In these panels, blue and yellow points represent Al and N plasmas, respectively, and
green curves show the electron bulk velocity.
Panels of the fourth and fifth rows show the transverse magnetic field strength $B_y$ and
electron density $n_e$, respectively.  
The bottom panels represent electron temperature $T_e$ (blue line)
and emissivity of the plasma self-emission $n_e^2/\sqrt{T_e}$ normalized by far upstream value (red line).
Note that the electron density $n_e$ is displayed with a logarithmic scale.
\label{fig:PIC_3.5T}
}
\end{figure*}

The results at $t\approx5$, 10, and 15~ns are shown in Fig.~\ref{fig:PIC_3.5T}.
After $t=10$~ns, incoming N ions are reflected at the region of strong magnetic field
that is the compressed external field (center and right panels of Fig.~\ref{fig:PIC_3.5T}).
The reflection point is then the upstream side of the interface between Al and N plasmas 
(structure ``P2'' shown in Fig.~\ref{fig:PIC_3.5T}, and in Figs.~\ref{fig:sop}(c) and \ref{fig:extmag}(c)).
Hence, after $t=10$~ns, incoming N ions are reflected in the N plasma.
These reflected N ions form the structure ``R2'' (see  Figs.~\ref{fig:sop}(c), \ref{fig:extmag}(a), and \ref{fig:extmag}(c)).
The N ions reflected at $t\approx10$~ns are responsible for  ``R2'' at $t=23$~ns.
In earlier epoch ($t\approx5$~ns: left panels of Fig.~\ref{fig:PIC_3.5T}), incoming N ions penetrates into Al plasma and they are reflected by the Biermann field in the Al plasma 
(left panels of Fig.~\ref{fig:PIC_3.5T}).
However, such ions have already come back to N plasma before $t=23$~ns, and have turned again by the external field, going back
toward the downstream region (it has been passed 0.7 times the ion gyro period). Hence, they are not at the edge R2 when $t=23$~ns,
but at around $X'=0.9$--1.0~cm.

\section{Effects of Coulomb collisions}\label{app:coulomb}

In this paper, we have mainly considered collisionless processes.
However, collisional effects are not negligible in some cases.
Our current 1D PIC simulations do not capture the role of Coulomb collisions.
The mean-free path of electron-electron Coulomb collision is smaller than any other scales in most cases with typical parameters, 
so that electrons are hydrodynamically coupled.
Below, we discuss effects of ion-ion collisions
for structures P1, P2, R1, and R2.
In the following, we adopt an approximate formula of the mean-free path of ions with mass $m_1$, charge $Z_1e$,
and initial velocity $v_1$, running into
the plasma with ion mass $m_2$, charge $Z_2e$, electron density $n_{e2}$, ion density $n_{i2}$,
electron temperature $T_{e2}$, and ion temperature $T_{i2}$,
which is given by
$\lambda_{ii}=\mu m_1 v_r{}^4/4\pi n_{2i}Z_1{}^2Z_2{}^2e^4\ln\Lambda$,
where $\mu=m_1m_2/(m_1+m_2)$ and the mean relative velocity 
$v_r=\sqrt{v_1{}^2+(8kT_{i2}/\pi m_2)}$.
The Coulomb logarithm is calculated with $\Lambda=\min\{\lambda_{e2},\lambda_{i2}\}/b_L$,
where $b_L=Z_1Z_2e^2/\mu v_r$, and
$\lambda_{e2}=\sqrt{kT_{e2}/4\pi n_{e2}e^2}$ and $\lambda_{i2}=\sqrt{kT_{i2}/4\pi Z_2n_{e2}e^2}$
are electron and ion Debye lengths, respectively \cite{Spitzer1965,Pauls1995}.

\subsection{Structure P1 and P2}\label{app:coulombP1P2}

Our claim is that N ions penetrating into Al plasma is reflected by Biermann magnetic field in the Al plasma.
Although the Al plasma parameters are uncertain at times and positions of our interest, 
we set $T_e=T_i=100$~eV, $Z_{\rm Al}=9$,
 the ion density $n_{\rm Al}=1\times10^{18}$cm$^{-3}$, and the magnetic field strength $B_{\rm Al}=10$~T
 as typical values.
A trajectory of P1 and P2, $X'(t)$, gives the velocity, $dX'/dt$, of 510~km~s$^{-1}$ at $t=10$~ns,
and of 406~km~s$^{-1}$ at $t=23$~ns, so we set the relative velocity $v_r=500$~km~s$^{-1}$ 
as a fiducial value.
The mean-free path of N ions with $Z_{\rm N}=3$ in the Al plasma is then
%
$\lambda_{ii}\approx0.5~{\rm cm}\,
(3/Z_{\rm N})^{2}
(9/Z_{\rm Al})^{2}
(8.7/\ln\Lambda)
v_{r,500}{}^4
/n_{\rm Al,18}$,
and gyro radius of the N ions is given by
$r_{g,{\rm N}}=0.24~{\rm cm}\,
(3/Z_{\rm N})
v_{r,500}
(B_{\rm Al}/10~{\rm T})^{-1}$,
where $v_{r,500}=v_r/500~{\rm km}~{\rm s}^{-1}$ and
$n_{\rm Al,18}=n_{\rm Al}/10^{18}{\rm cm}^{-3}$, so that
their ratio is
 $\lambda_{ii}/r_{g,{\rm N}}=2.2
 (3/Z_{\rm N})
 (9/Z_{\rm Al})^{2}
(8.7/\ln\Lambda)
(B_{\rm Al}/10~{\rm T})
v_{r,500}{}^3
/n_{\rm Al,18}$.
When we only change temperatures to $T_e=T_i=10$~eV, then we get
$\lambda_{ii}/r_{g,{\rm N}}=2.6$.
If $Z_{\rm N}=6$ with the other parameters being fiducial, the ratio becomes $\lambda_{ii}/r_{g,{\rm N}}=1.2$.
In these cases, although Coulomb collision is marginally subdominant,  it is non-negligible.
However, at least, it can be said that the Biermann field plays a crucial role in N ion reflection in the Al plasma.
Note that while the electron-electron Coulomb collision mean-free path is much smaller than 
$\lambda_{ii}$ and $r_{g,{\rm N}}$, electrons around P1 and P2 cannot generate an electric field that is able to reflect
incoming N ions.

\subsection{Structure R1}\label{app:coulombR1}

An explanation of structure R1 is fast Al ions penetrating into initial upstream N plasma
as shown in our 1D PIC simulation.
The mean-free path of the Al ions with $Z_{\rm Al}=9$ in the N plasma 
($T_e=T_i=6$~eV, $Z_{\rm N}=3$, and the N ion density $n_{\rm N}=3.2\times10^{17}$cm$^{-3}$ corresponding
to $P_N=5$~Torr) is estimated as
$\lambda_{ii}\approx11~{\rm cm}\,
(3/Z_{\rm N})^{2}
(9/Z_{\rm Al})^{2}
(8.1/\ln\Lambda)
v_{r,700}{}^4
$, where $v_{r,700}=v_r/700~{\rm km}~{\rm s}^{-1}$.
If R1 is the head of N ions reflected by interface P1, their mean-free path 
is also large ($\lambda_{ii}\gtrsim10v_{r,700}{}^4$~cm for reflected ion charge number
$Z_{\rm N}=3$--6).
Hence, the collisional effect is negligible for the propagation of R1.

\subsection{Structure R2}\label{app:coulombR2}

Our claim is that R2 at $t=23$~ns is the edge of N ions that are reflected
by compressed external magnetic field in the N plasma just upstream of P2 at $t\approx10$~ns.
Although the parameters of the N plasma at the reflection region are again uncertain,
we set $T_e=200$~eV, $T_i=30$~eV, the electron density $n_e=1\times10^{19}$cm$^{-3}$,
and the strength of the compressed external field $B_N=10$~T, considering the result of our 
1D PIC simulation at $t=10$~ns.
The value of $Z_{\rm N}$ is also uncertain but for the worst case, we assume
that both incoming N ions and the N plasma have $Z_{\rm N}=6$.
Then we obtain the ion-ion mean-free path of incoming N ions
$\lambda_{ii}\approx0.2~{\rm cm}\,
(6/Z_{\rm N})^{3}
(7.5/\ln\Lambda)
v_{r,500}{}^4
/n_{e,19}$
and their gyro radius
$r_{g,{\rm N}}=0.12~{\rm cm}\,
(6/Z_{\rm N})
v_{r,500}
(B_{\rm N}/10~{\rm T})^{-1}$, where $v_{r,500}=v_r/500~{\rm km}~{\rm s}^{-1}$ and
$n_{e,19}=n_{e}/10^{19}{\rm cm}^{-3}$.
Hence, the ratio of these scale length is
$\lambda_{ii}/r_{g,{\rm N}}\approx1.3
(3/Z_{\rm N})^2
(7.5/\ln\Lambda)
(B_{\rm Al}/10~{\rm T})
v_{r,500}{}^3
/n_{e,19}$.
According to our PIC simulation results, the electron density $n_e\lesssim10^{19}$cm$^{-1}$ and
the field strength of the compressed external magnetic field $B_N\gtrsim10$~T at the reflection region.
Furthermore, the incoming N ions have $Z_N=3$.
Hence, we expect that the ratio $\lambda_{ii}/r_{g,{\rm N}}$ should be larger, and
ion-ion collision is subdominant for the N ion reflection for the origin of R2 at $t=23$~ns.

After the reflection at $t\approx10$~ns, the reflected N ions return to 
the upstream N plasma ($T_e=T_i=6$~eV, $Z_{\rm N}=3$, and the N ion density $n_{\rm N}=3.2\times10^{17}$cm$^{-3}$)
and gyrate due to the external magnetic field.
Assuming that the reflected ions have $Z_{\rm N}=6$ with velocity $400$~km~s$^{-1}$, 
which are values inferred from the TS analysis results at $t=23$~ns, we get
 the ion-ion mean-free path of reflected N ions
$\lambda_{ii}\approx1.2~{\rm cm}$.
Just after the reflection ($t\approx10$~ns), the velocity of the reflected ions should be larger than $400$~km~s$^{-1}$ as shown in 1D
PIC simulation, so that the Coulomb collision is negligible when reflected ions move in the upstream N plasma.

Coulomb collision also changes ionization state of N ions, that is the value of $Z_{\rm N}$.
If the N plasma with $T_e\approx T_i\gtrsim200$~eV is in collision-ionization equilibrium 
(as assumed in analyzing TS IAW data),
then $Z_{\rm N}$ becomes large, e.g., $Z_{\rm N}=5$--7.
However, the equilibrium state is achieved when $n_et_p\gtrsim10^{11-12}$s~cm$^{-3}$,
where $t_p$ is the plasma age, that is, the elapsed time from the plasma generation  \cite{Spitzer1965}.
The value of $n_et_p$ to reach the equilibrium depends on the initial values of $T_e$, $T_i$, $n_e$, and $Z_{\rm N}$.
In our case, $t_p$ is roughly given by the crossing time of the high-temperature reflection region,
which has typical width of $0.1$~cm, so that
$t_p\approx2v_{r,500}^{-1}$~ns and $n_et_p\approx2\times10^{10}n_{e,19}v_{r,500}^{-1}$~s~cm$^{-3}$.
The time evolution of $Z_{\rm N}$ should be calculated, but it is complicated.
Such detailed calculation is beyond the scope of this paper.
However, for our values of $n_et_p$, the value of  $Z_{\rm N}$ may be smaller than the value 
for the equilibrium state ($Z_{\rm N}=5$--7).




\begin{acknowledgments}
The authors would like to acknowledge the dedicated technical
support of the staff at the Gekko-XII facility for the laser operation,
target fabrication, and plasma diagnostics.
We thank A.~Takamine and H.~Maeda for useful discussions.
We would also like to thank anonymous referees for their careful reading and helpful comments to improve the paper.
This research was partially
supported by 
the joint research project of Institute of Laser Engineering, Osaka University, 
JSPS KAKENHI Grants No. 
%
18H01232, 19H01868, 17K18270, 
18H01245, 18H01246, 19K14712, 17J03893, and
15H02154, 
the Sumitomo Foundation for environmental research projects (203099) (S.M.),
and
JSPS Core-to-Core Program B: Asia-Africa Science Plat-forms Grant No. JPJSCCB20190003 (Y.~Sakawa).
Computer simulations were performed on the CIDAS
supercomputer system at the Institute for Space-Earth Environmental
Research in Nagoya University under the joint research
program.
R.Y.  and S.J.T. deeply appreciate Aoyama Gakuin University Research Institute for partially funding our research.

\end{acknowledgments}



\bibliography{paper}


\end{document}